\documentclass[twocolumn]{aastex631}

\received{November 29, 2024}
\revised{June 23, 2025}
\accepted{June 24, 2025}

\usepackage{graphicx}	
\usepackage{amsmath}	
\usepackage{xcolor}
\usepackage{ragged2e}
\usepackage{hyperref}

\shorttitle{Multi-phase ISM modeling of an LBG at $z=8.312$}
\shortauthors{Hagimoto et al.}

\begin{document}

\title{Compact Ionized Gas Region Surrounded by Porous Neutral Gas in a Dusty Lyman Break Galaxy at Redshift $z=8.312$}

\correspondingauthor{Masato Hagimoto}
\email{hagimoto@a.phys.nagoya-u.ac.jp}

\author[0000-0001-8083-5814]{Masato Hagimoto}
\affiliation{Department of Physics, Graduate School of Science, Nagoya University, Furo, Chikusa, Nagoya, Aichi 464-8602, Japan}

\author[0000-0003-4807-8117]{Yoichi Tamura}
\affiliation{Department of Physics, Graduate School of Science, Nagoya University, Furo, Chikusa, Nagoya, Aichi 464-8602, Japan}

\author[0000-0002-7779-8677]{Akio K. Inoue}
\affiliation{Department of Physics, School of Advanced Science and Engineering, Faculty of Science and Engineering, Waseda University, 3-4-1, Okubo, Shinjuku, Tokyo 169-8555, Japan}
\affiliation{Waseda Research Institute for Science and Engineering, Faculty of Science and Engineering, Waseda University, 3-4-1 Okubo, Shinjuku, Tokyo 169-8555, Japan}

\author[0000-0003-1937-0573]{Hideki Umehata}
\affiliation{Institute for Advanced Research, Nagoya University, Furo, Chikusa, Nagoya 464-8602, Japan}
\affiliation{Department of Physics, Graduate School of Science, Nagoya University, Furo, Chikusa, Nagoya, Aichi 464-8602, Japan}

\author[0000-0002-5268-2221]{Tom J. L. C. Bakx}
\affiliation{Department of Space, Earth, \& Environment, Chalmers University of Technology, Chalmersplatsen 4 412 96 Gothenburg, Sweden}

\author[0000-0002-0898-4038]{Takuya Hashimoto}
\affiliation{Graduate School of Pure and Applied Sciences, University of Tsukuba, 1-1-1 Tennodai, Tsukuba, Ibaraki 305-8571, Japan}
\affiliation{Tomonaga Center for the History of the Universe, University of Tsukuba, 1-1-1 Tennodai, Tsukuba, Ibaraki 305-8571, Japan}

\author[0000-0003-4985-0201]{Ken Mawatari}
\affil{Waseda Research Institute for Science and Engineering, Faculty of Science and Engineering, Waseda University, 3-4-1 Okubo, Shinjuku, Tokyo 169-8555, Japan}

\author[0000-0001-6958-7856]{Yuma Sugahara}
\affiliation{Department of Physics, School of Advanced Science and Engineering, Faculty of Science and Engineering, Waseda University, 3-4-1, Okubo, Shinjuku, Tokyo 169-8555, Japan}

\author[0000-0001-7440-8832]{Yoshinobu Fudamoto}
\affiliation{Center for Frontier Science, Chiba University, 1-33 Yayoi-cho, Inage-ku, Chiba 263-8522, Japan}

\author[0000-0002-6047-430X]{Yuichi Harikane}
\affiliation{Institute for Cosmic Ray Research, The University of Tokyo, 5-1-5 Kashiwanoha, Kashiwa, Chiba 277-8582, Japan}

\author[0000-0003-3278-2484]{Hiroshi Matsuo}
\affiliation{National Astronomical Observatory of Japan, 2-21-1 Osawa, Mitaka, Tokyo 181-8588, Japan}
\affiliation{The Graduate University for Advanced Studies (SOKENDAI), 2-21-1 Osawa, Mitaka, Tokyo 181-8588, Japan}

\author[0000-0002-9695-6183]{Akio Taniguchi}
\affiliation{Kitami Institute of Technology, 165 Koen-cho, Kitami, Hokkaido 090-8507, Japan}
\affiliation{Department of Physics, Graduate School of Science, Nagoya University, Furo, Chikusa, Nagoya, Aichi 464-8602, Japan}

\begin{abstract}

Porous interstellar medium (ISM) structure in galaxies at the epoch of reionization (EoR) gives us a hint to understand what types of galaxies contribute to reionization. 
Although recent studies have pointed out the positive correlation between high ionizing photon escape fractions and high \textsc{[O\,iii]}~$88~\mu\mathrm{m}$-to-\textsc{[C\,ii]}~$158~\mu\mathrm{m}$ ratios found in UV-luminous star-forming galaxies at $z > 6$ with ALMA, previous studies have paid little attention to the neutral gas porosity that allows ionizing photons to escape.
Here, we present a detailed analysis of a $z=8.312$ Lyman break galaxy, MACS0416\_Y1 with a high $L_\mathrm{[OIII]88}/L_\mathrm{[CII]158}$ ratio ($\approx9$) and dust continuum detection. 
We construct a multi-phase ISM model incorporating the neutral gas covering fraction ($cov_\mathrm{PDR}$). 
The best-fit model reveals a $cov_\mathrm{PDR}\approx25 \%$, indicating that $\approx75 \%$ of the ionized gas region is exposed to intercloud space. 
We confirm that our conclusions hold even when varying star-formation history, stellar age, gas/stellar metallicity, and carbon-to-oxygen abundance ratio.
This finding meets one of the necessary conditions for galaxies to have a non-zero escape fraction of ionizing photons and supports recent studies that galaxies with a high \textsc{[O\,iii]}~$88~\mu\mathrm{m}/$\textsc{[C\,ii]}~$158~\mu\mathrm{m}$ ratio, such as MACS0416\_Y1, could contribute to cosmic reionization.
Furthermore, the modeled \textsc{H\,ii} region with the best-fitting parameters has a typical size ($D=0.90~\mathrm{pc}$) and gas density ($\log n_\mathrm{H,c}/\mathrm{cm^{-3}}=2.7$) that are comparable to local compact \textsc{H\,ii} regions.
This suggests that the \textsc{H\,ii} regions in MACS0416\_Y1 are in an early evolutionary stage. 
\end{abstract}


\section{Introduction} \label{sec:intro}
Chemically-enriched, massive galaxy populations at the epoch of reionization (EoR) play a significant role in galaxy formation and evolution through active star formation and feedback \citep[e.g.,][]{Behroozi2019}.
They offer unexpected insights into the early Universe, including the presence of abundant luminous galaxy populations at $z\gtrsim10$ \citep[e.g., ][]{Bunker2023,Curtis-Lake2023,Harikane2024,Zavala2024,Castellano2024,Carniani2024} and over-density regions serving as sites of metal-enrichment \citep[e.g., ][]{Morishita2023,Hashimoto2023,Jones2024_HFLS3}.
These galaxies could also be dominant contributors to cosmic reionization under the late ionization scenario proposed by \citet{Naidu2020}.
This scenario has recently been supported by statistical analysis of Lyman-$\alpha$ damping wings \citep{Umeda2023} and the velocity offset of Lyman-$\alpha$ emission \citep{Nakane2024} based on spectroscopic data from the James Webb Space Telescope (JWST).

The neutral gas phase, specifically the photo-dissociation region \citep[PDR;][]{Hollenbach1999,Wolfire2022}, and/or the molecular gas phase are crucial for understanding the physical properties of the interstellar medium (ISM), including the escape of ionizing photons into intergalactic space.
JWST observations of rest-frame ultraviolet (UV)/optical emission lines are not appropriate to reveal the physical properties of these gas phases since atoms/ions that radiate bright lines to be detectable even in high-$z$ galaxies typically have higher ionization potentials than hydrogen ($13.6~\mathrm{eV}$)\footnote{The ionization potential of sulfur ($10.36~\mathrm{eV}$) is lower than that of hydrogen, but \textsc{[S\,ii]} lines are not bright in PDR because collisions with electrons emit the lines \citep{Draine2011}.}.
The rest-frame far-infrared (FIR) \textsc{[C\,ii]}$~158~\mathrm{\mu m}$ emission line and dust continuum observations with the Atacama Large Millimeter/submillimeter Array (ALMA) offer complementary insights \citep[e.g.,][]{Hashimoto2023,Fujimoto2024}.
Dust grains absorb UV radiation from ionizing sources and re-emit thermal continuum emission, making rest-frame FIR dust continuum emission a reliable tracer of dust-obscured star formation \citep[e.g.,][]{Madau2014,Fudamoto2021,Zavala2021,Algera2023} and dust mass \citep[e.g.,][]{Tamura2019,Bakx2021,Sugahara2021,Witstok2022,Algera2024} in galaxies.
The \textsc{[C\,ii]}$~158~\mathrm{\mu m}$ is one of the primary coolants of the ISM in the FIR emission lines in the local Universe \citep[e.g.,][]{Madden2013,Diaz-Santos2017,Herrera-Camus2018} and is widely used as a good tracer of global star-formation \citep[e.g.,][]{DeLooze2011,DeLooze2014,Herrera-Camus2015} and the PDR \citep{Diaz-Santos2017,Cormier2019}.
Taking advantage of its luminous nature, ALMA has routinely observed redshifted \textsc{[C\,ii]}$~158~\mathrm{\mu m}$ toward galaxies at $z\geq4.5$, including four ALMA large programs (ALPINE; \citealt{LeFerve2020}, REBELS; \citealt{Bouwens2022}, CRISTAL; \citealt{Herrera-Camus2025}, ASPIRE\footnote{\url{https://aspire-quasar.github.io/project-alma.html}}).
These observations reveal that this emission line remains a good tracer of global star-formation rates (SFRs) even at such high redshifts \citep[e.g.,][]{Schaerer2020,Liang2024}.
In addition, \citet{Witstok2022} reported a galaxy at $z\sim7$ that \textsc{[C\,ii]}$~158~\mathrm{\mu m}$ predominantly originates from PDRs based on the comparison between the \textsc{[C\,ii]} and \textsc{[N\,ii]}~$205~\mu\mathrm{m}$ emission lines, which have similar critical densities but are emitted by ions with different ionization potentials.

The \textsc{[O\,iii]}$~88~\mathrm{\mu m}$ line is known to be another brightest emission line in the local Universe \citep[e.g.,][]{Takami1987,Mizutani2002,Matuo2009,Cormier2015,Herrera-Camus2018}.
It is considered to trace ionized gas regions because of high ionization potential ($35.1~\mathrm{eV}$).
Since \citet{Inoue2016} reported a detection of redshifted \textsc{[O\,iii]}$~88~\mathrm{\mu m}$ emission in a Lyman-$\alpha$ emitter (LAE) at $z=7.21$ with ALMA, based on the model predictions in \citet{Inoue2014}, subsequent studies have detected this line in UV luminous star-forming galaxies \citep[LAEs and Lyman break galaxies, LBGs; e.g.,][]{Hashimoto2018,Hashimoto2019BTD,Tamura2019,Harikane2020,Akins2022,Witstok2022,Fujimoto2024}, dusty star-forming galaxies \citep[e.g.,][]{Marrone2018,Tadaki2022,Algera2024,Bakx2024}, and quasars \citep[e.g.,][]{Walter2018,Hashimoto2019QSO} at $z\gtrsim6$.
The luminosity ratios of \textsc{[O\,iii]}$~88~\mathrm{\mu m}$ to \textsc{[C\,ii]}$~158~\mathrm{\mu m}$ for UV luminous star-forming galaxies at $z\gtrsim6$ have been found to be $L_\mathrm{[OIII]}/L_\mathrm{[CII]}\gtrsim3$ \citep{Inoue2016,Laporte2019,Hashimoto2019BTD,Bakx2020CII,Carniani2020,Harikane2020,Witstok2022}, which is higher than that observed in more massive and dusty galaxies at the same epoch \citep[$L_\mathrm{[OIII]}/L_\mathrm{[CII]}\lesssim1.5$; e.g.,][]{Marrone2018,Algera2024} and even in local low-metallicity dwarfs \citep[$L_\mathrm{[OIII]}/L_\mathrm{[CII]}\approx2$;][]{Cormier2015} with the exception of Pox 186 with a \textsc{[O\,iii]}$~88~\mathrm{\mu m}/$\textsc{[C\,ii]}$~158~\mathrm{\mu m}$ ratio of $\gtrsim10$ \citep{Kumari2024MNRAS}.

The origins of such high $L_\mathrm{[OIII]}/L_\mathrm{[CII]}$ ratios in galaxies at $z\gtrsim6$ have been explored through both observational and numerical studies.
Possible origins include the extraordinary nature of ionizing sources (high-ionization parameters, bursty star-formation histories, and top-heavy initial mass functions) and/or gas physical and chemical conditions (low carbon-to-oxygen abundance ratios, low covering fractions of a PDR surrounding the ionized gas region, and high Lyman continuum escape fractions) although observational biases, such as strong cosmic microwave background attenuation, spatially extended \textsc{[C\,ii]} haloes, inclination effects, might be the case \citep{Arata2020,Bakx2020CII,Carniani2020,Harikane2020,Vallini2021,Katz2022,Ramambason2022,Sugahara2022,Ura2023,Bakx2024,Fujimoto2024,Nyhagen2024,Schimek2024}.

The key to resolving the degeneracy between ionizing sources and gas physical conditions lies in performing more detailed modeling of the ISM properties.
Notably, \citet{Cormier2019} (and also \citealt{Ramambason2022}) developed a multi-phase ISM model incorporating neutral gas ``porosity'' using photo-ionization code, \textsc{Cloudy} \citep{Ferland2017}.
They estimated the ionization parameters, gas densities, and neutral gas porosity for the Herschel dwarf galaxy survey samples \citep{Madden2013} based on FIR emission lines and dust continuum. 
They show that galaxies with lower metal abundances tend to have larger neutral gas porosity, suggesting that low metal dwarfs in the early Universe might have porous neutral gas and contribute to cosmic reionization.
In general, applying their method to high-redshift galaxies is a challenging task because of the poorer observational constraints; however, for certain galaxies with detections of at least two FIR emission lines and dust continuum, it is possible to investigate these three parameters, i.e., the ionization parameter, the gas density, and the neutral gas porosity.

In this paper, we investigate the physical properties of the ISM in MACS0416\_Y1 at $z=8.312$, one of the LBGs with ALMA detections in \textsc{[O\,iii]}~$88~\mathrm{\mu m}$, \textsc{[C\,ii]}~$158~\mathrm{\mu m}$, and dust continuum, by comparing observations and photoionization models.
We briefly describe the target in Section~\ref{sec:target}, and explain our models in Section~\ref{sec:method}. 
We show the results for the modeling and discuss consistency with other observations in Section~\ref{sec:result}. 
In Section~\ref{sec:oldstellarpopulation}, we discuss a potential contribution of old stellar populations to the young ones from the spectral energy distribution (SED), and Section~\ref{sec:size_HIIregion} presents the physical properties of the modeled \textsc{H\,ii} region with the best-fit parameters.
Section~\ref{sec:systematics} discusses the systematic differences related to parameter selection in our model.
Finally, we provide a summary of our results in Section~\ref{sec:summary}. 
Throughout this paper, we adopted a spatially flat $\Lambda$CDM cosmology with the best-fit parameters derived from the Planck results \citep{Planck2018parameters}, which are $\Omega_\mathrm{m} = 0.310$, $\Omega_\mathrm{\Lambda} = 0.690$ and $h = 0.677$. 

\section{Target: MACS0416\_Y1} \label{sec:target}
Our target, MACS0416\_Y1 (hereafter Y1), was firstly found as a bright, $Y$-band dropout LBG \citep{Infante2015,Laporte2015} and is moderately gravitationally magnified ($\mu_\mathrm{g}=1.43\pm0.04$\footnote{Recently two different groups reported revised estimates based on the new data, but they are completely different ($\mu_\mathrm{g}=1.21$ in \citealt{Ma2024}, and $\mu_\mathrm{g}=1.60^{+0.01}_{-0.02}$ in \citealt{Harshan2024}). In this paper, we use the same factor to correct all luminosities, so which value is used does not affect the result.}; \citealt{Kawamata2016}) by the foreground massive galaxy cluster, MACS0416.1$-$2403.
Follow-up observations with ALMA identified the spectroscopic redshift to be $z=8.312$ through the detection of the rest-frame FIR \textsc{[O\,iii]}$~88~\mathrm{\mu m}$ and \textsc{[C\,ii]}$~158~\mathrm{\mu m}$ emission lines \citep{Tamura2019,Bakx2020CII}.
The global \textsc{[O\,iii]}$~88~\mathrm{\mu m}$-to-\textsc{[C\,ii]}$~158~\mathrm{\mu m}$ luminosity ratio is $\approx9.3\pm2.6$ \citep{Bakx2020CII}, making it one of the galaxies with the highest luminosity ratios in the EoR. 
Even when correcting for the surface brightness dimming effect, the ratio remains high \citep[$\approx8\pm2$,][]{Carniani2020}.

The rest-frame $90~\mathrm{\mu m}$ dust continuum emission was also detected \citep{Tamura2019} while it was neither detected at rest-frame $120~\mathrm{\mu m}$ nor $160~\mathrm{\mu m}$ \citep{Bakx2020CII}.
This suggests that the SED within the range of $90 \leq \lambda_\mathrm{rest}/\mathrm{\mu m}\leq 160$ is close to the Rayleigh-Jeans regime with a high dust temperature of $T_\mathrm{d}\gtrsim80~\mathrm{K}$ \citep{Bakx2020CII,Sommovigo2022,Fudamoto2023,Algera2024,Jones2024_Y1}.
Such a high $T_\mathrm{d}$ leads to a high infrared luminosity ($L_\mathrm{IR}\approx1\times10^{12}~\mathrm{L_\odot}$ under $T_\mathrm{d}=80~\mathrm{K}$ and a dust emissivity index $\beta_\mathrm{d}=2$), indicating that dust clouds are exposed to intense and hard UV radiation from young, compact star-forming regions.
These conditions potentially result in the high-\textsc{[O\,iii]}-to-\textsc{[C\,ii]} ratios observed in this galaxy.

\citet{Tamura2023} presented the 300~pc scale high-resolution imaging of the  \textsc{[O\,iii]}$~88~\mathrm{\mu m}$ emission line and rest-frame $90~\mathrm{\mu m}$ dust continuum emission.
They compared the spatial distributions of \textsc{[O\,iii]}$~88~\mathrm{\mu m}$, dust continuum, and young stellar populations traced by the HST/F160W filter (rest-frame $\sim1600~$\AA).
They revealed that the \textsc{[O\,iii]}$~88~\mathrm{\mu m}$ emission has three peaks likely associated with three young stellar components, while the dust continuum has two peaks distributed between the stellar clumps.
Such spatial segregation on the sub-kpc scale between young stellar and dust clumps implies ionizing photons could escape into the intercloud space through channels of low dust attenuation regions.
In addition, new high-resolution \textsc{[C\,ii]} imaging has revealed multiple \textsc{[C\,ii]} clumps in the velocity regime (Bakx et al., in prep.).

Moreover, JWST observations have recently revealed the rest-frame optical view of Y1. 
\citet{Desprez2024} presented the NIRCam images of eight bands taken as part of the CAnadian NIRISS Unbiased Cluster Survey \citep[CANUCS,][]{Willott2022}. 
They performed the SED fitting for the entire galaxy system using those data and archival HST/WFC3 data with \texttt{Dense Basis} \citep{Iyer2019} and \texttt{Bagpipes} \citep{Carnall2018}, and estimated the stellar masses to be $\log(M_\star)\approx8.5$--$9.0$. 

\citet{Ma2024} also showed NIRCam images of eight filters taken by Prime Extragalactic Areas for Reionization and Lensing Science \citep[PEARLS;][]{Windhorst2023}.
Utilizing these NIRCam data and HST/WFC3 data obtained by the Hubble Frontier Fields program, they performed the SED fitting with \texttt{Bagpipes} for the entire galaxy system.
As a result, they found that Y1 is young ($\mathrm{age}\approx4.7~\mathrm{Myr}$) and in a starburst phase ($\mathrm{SFR}\approx160~\mathrm{M_\odot~yr^{-1}}$). 
It has a similar stellar mass of $M_\star\approx7.7\times10^8~\mathrm{M_\odot}$ to that obtained by \citet{Desprez2024} and a flat UV slope ($-1.8\lesssim\beta_\mathrm{UV}\lesssim-1.0$) because of the dust extinction ($0.92\lesssim A_\mathrm{V}\lesssim1.10$). 
In addition, they decomposed three UV-bright clumps and performed the SED fit to these three individual clumps, which showed similar stellar masses ($M_\star\sim3.3\times10^8~\mathrm{M_\odot}$) but were relatively diverse in other properties ($1.6\lesssim \mathrm{Age/Myr}\lesssim10$, $30\lesssim \mathrm{SFR/\mathrm{M_\odot~yr^{-1}}}\lesssim200$, $0.6\lesssim A_\mathrm{V}\lesssim1.2$, and $-2.5\lesssim\beta_\mathrm{UV}\lesssim-1.0$).
Moreover, they reported detection of the \textsc{[O\,iii]}$\lambda\lambda4960+5008$ and \textsc{[O\,ii]}$\lambda\lambda3727+3729$ emission lines with NIRCam wide-field slitless spectroscopy \citep[PID 3538; ][]{Iani2023}.

\begin{figure}
\includegraphics[width=0.45\textwidth]{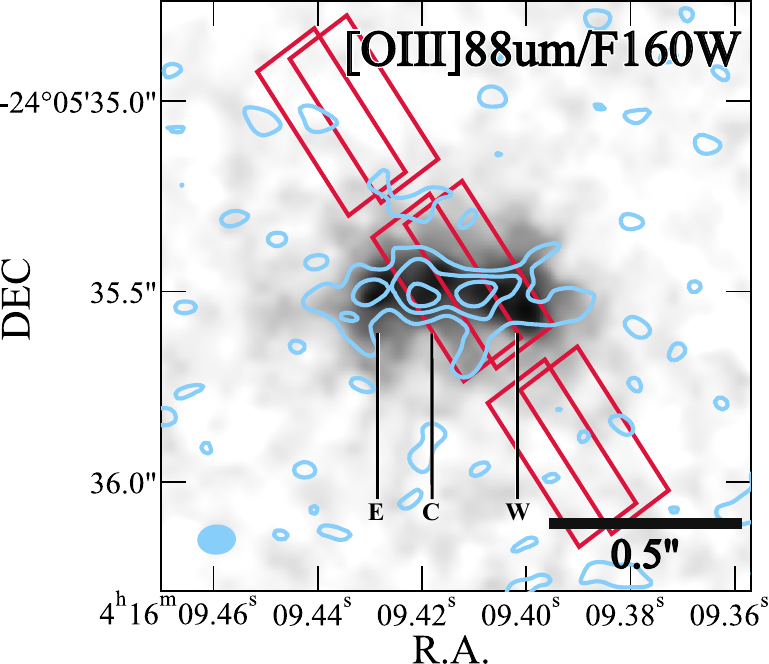}
\caption{Rest-frame UV images (HST/F160W) with the contour of  \textsc{[O\,iii]}$~88~\mathrm{\mu m}$ moment 0 map reported in \citet{Tamura2023}. The contours are $(-2,\, 2,\, 4,\, 6,\, 8)\times\sigma$ where $1\sigma=11~\mathrm{mJy~km~s^{-1}~beam^{-1}}$. We also overlaid the two slits used to obtain the spectrum with JWST/NIRSpec in \citet{Harshan2024}. These two slits do not cover the eastern UV clump (`E' in the figure) as they reported. We find that the eastern peak of \textsc{[O\,iii]}$~88~\mathrm{\mu m}$ image is not covered by two slits as well. These findings suggest that their spectrum can underestimate the flux of the entire system by $\approx2/3$.
}
\label{fig:Y1images}
\end{figure}

\citet{Harshan2024} presented a NIRSpec micro-shutter assembly (MSA) spectrum observed as a part of the CANUCS, in which a total of eight emission lines, including blended ones (\textsc{[O\,ii]}$\lambda\lambda3727+3729$ and $\mathrm{H}\gamma+$\textsc{[O\,iii]}$\lambda4364$), are evident.
They predicted physical parameters, such as electron temperature ($T_\mathrm{e}$), gas-phase metallicity ($Z_\mathrm{gas}$), the ionization parameter ($U$), and electron density ($n_\mathrm{e}$) by using obtained line ratios and \texttt{PyNeb} \citep{Luridiana2015} or empirical relations between line ratios and physical parameters provided by \citet{Izotov2006, Kewley2019, Nicholles2020}.
They found that Y1 has comparable $T_\mathrm{e}$ ($\approx18000~\mathrm{K}$) but is a more chemically evolved system ($Z_\mathrm{gas}\sim15\%~\mathrm{Z_\odot}$, $\log U\approx-2.5$) than other UV-bright galaxies at $z\sim6$--$8$, which are confirmed by NIRSpec spectroscopy \citep[e.g.,][]{Schaerer2022,Cameron2023}.
Electron density was measured to be $\sim100~\mathrm{cm^{-3}}$ with the line ratio of \textsc{[O\,iii]}$~88~\mathrm{\mu m}$ and \textsc{[O\,iii]}$\lambda5008$.
A caveat includes that their results are based on the data taken within the two shutters covering only two of three UV clumps, indicating they do not represent the total galaxy system (see Figure~\ref{fig:Y1images}). 

\section{Methodology} \label{sec:method}
\subsection{Model parameters} \label{sec:param}
We used the photoionization code \textsc{Cloudy} version 17.03 \citep{Ferland2017} to construct the multi-phase ISM model of Y1.
We assume that the ISM consists of an ionized (\textsc{H\,ii} region) and neutral gas (PDR) region \citep[e.g.,][]{Cormier2019}. 
The ISM is assumed to be spherically symmetric, with a central ionizing source completely surrounded by ionized gas. 
Our goal is to estimate the physical properties of the typical ISM in this galaxy by comparing our model with observations, where `typical' implies that this galaxy comprises numerous regions of such medium.
In what follows, we explain the main input parameters for our fiducial model, which is also summarized in Table~\ref{tab:inputparameters}.

\begin{table}
\caption{Input parameters for our fiducial model.}
\begin{center}
\scalebox{0.85}{
    \begin{tabular*}{9.2cm}{ll}
        \hline
        \hline
        \multicolumn{2}{c}{Fixed parameters} \\
        \hline
        Stellar population & BPASS ver 2.0 \citep{Stanway2016} \\
        & Broken power-law IMF ($f(m)\propto m^\alpha$) \\
        & $\alpha=-1.30$ for $0.1 \leq M_\mathrm{star}/\mathrm{M_\odot} \leq 0.5$ \\
        & $\alpha=-2.35$ for $0.5 < M_\mathrm{star}/\mathrm{M_\odot} \leq 100$ \\
        & Continuous star formation for $4~\mathrm{Myr}$, \\
        & Stellar metallicity of $Z_\mathrm{star}=0.2~\mathrm{Z_\odot}$ \\
        & \citep{Tamura2019} \\
        & Including binary stars \\
        Background radiation & CMB at $z=8.3118$ \\
        & \citep{Tamura2019} \\
        & Cosmic ray background \\
        & \citep{Indriolo2007} \\
        Gas chemical & Solar one except for helium \\
        composition & \citep{Grevesse2010} \\
        & Equation~(\ref{eq:He-abandance}) for the helium abundance \\
        & \citep{Groves2004} \\
        Grains model &  ISM of Milky Way from \textsc{Cloudy} \\
        Gas-phase Metallicity & $Z_\mathrm{gas}=0.2~\mathrm{Z_\odot}(=Z_\mathrm{star})$ \\
        Density profile & $n_\mathrm{H}=n_\mathrm{H,c}\times\left(1+N(\mathrm{H})/10^{21}~\left[\mathrm{cm^{-2}}\right]\right)$ \\
        & \citep{Cormier2019} \\
        Stopping criteria & $\mathrm{H^+}/\mathrm{H_{tot}}=0.01$ for \textsc{H\,ii} region \\
        & \citep{Abel2005,Nagao2011} \\
        & $A_\mathrm{V}=5~\mathrm{mag}$ for PDR \\
        & \citep{Cormier2019} \\
        Dust temperature & $80~\mathrm{K} ^\mathrm{a}$ \\
        & \citep{Bakx2020CII} \\
        \hline
        \hline
        \multicolumn{2}{c}{Varied parameters} \\
        \hline
        $\log n_\mathrm{H,c}/\mathrm{cm}^{-3}$ & [1.0:4.0, 0.1 intervals]\\
        $\log U$ & [-4.0:0.0, 0.1 intervals] \\
        $cov_\mathrm{PDR}$$^\mathrm{b}$ & [0.0:1.0, 0.01 intervals] \\
    \hline
    \end{tabular*}
    }
\end{center}
\label{tab:inputparameters}
$^\mathrm{a}$ The dust temperature is not an input parameter of \textsc{Cloudy}, but this significantly affects our fitting result via infrared luminosity. We adopt the dust temperature of $80~\mathrm{K}$ in the fiducial model based on \citet{Bakx2020CII}.

$^\mathrm{b}$ $cov_\mathrm{PDR}$ is not an input parameter of \textsc{Cloudy}.
\end{table}

The incident spectrum consists of a stellar component and background radiation.
The stellar component is modeled using the Binary Population and Spectral Synthesis (BPASS) code version 2.0 \citep{Stanway2016,Eldridge2016}.
We employ a binary stellar model with the default initial mass function (IMF) in the BPASS code, which has a shape of $f(m)\propto m^\alpha$ ($m$ is the stellar mass) where the power-law indices, $\alpha$, are $\alpha = -1.30$ and $-2.35$ within the stellar mass ranges of $m = 0.1$--$0.5~\mathrm{M}_\odot$ and $0.5$--$100~\mathrm{M}_\odot$, respectively.
We assume a continuous star formation for $4~\mathrm{Myr}$ and a stellar metallicity of $0.2~\mathrm{Z_\odot}$, which were obtained by the rest-frame UV to FIR SED fits \citep{Tamura2019}. 
As background radiation, we adopt only the cosmic microwave background (CMB) at $z=8.3118$ \citep{Tamura2019}, and for cosmic ray background, we assume the default \textsc{Cloudy} assumptions \citep{Indriolo2007} as a first-order approach

We use the solar chemical composition for the gas-phase ISM \citep{Grevesse2010}.
Still, for helium abundance, we adopt 
\begin{equation}
    \mathrm{He/H}=0.0737+0.0293Z_\mathrm{gas}/\mathrm{Z_\odot}
    \label{eq:He-abandance}
\end{equation}
\citep{Groves2004,Sugahara2022} to account for the Big Bang nucleosynthesis and stellar yields.
We assume dust grains with the same size distribution and abundance pattern as the Milky Way, which is a default of \textsc{Cloudy}. 
The gas phase metallicity is assumed to be the same as the stellar one ($0.2~\mathrm{Z_\odot}$). 
We use the depth-dependent density profile following the previous study of the ISM modeling in low-metallicity environments \citep{Cormier2019,Ramambason2022}, given by 
\begin{equation}
    n_\mathrm{H}=n_\mathrm{H,c}\times\left(1+N(\mathrm{H})/10^{21}~\mathrm{cm^{-2}}\right),
    \label{eq:density-law}
\end{equation} 
where $n_\mathrm{H,c}$ is the hydrogen volume density at the gas-illuminated face, and $N(\mathrm{H})$ is the hydrogen column density at a given depth.
In this formulation, the density is roughly constant within the \textsc{H\,ii} region and increases along with hydrogen column density in the neutral gas (see also Figure~2 in \citealt{Cormier2019}). 
This represents the density profile as an intermediate choice between the two most common extreme regimes -- constant pressure and constant density. 
We make two models with different stopping criteria to obtain intensities from PDRs and only the \textsc{H\,ii} regions separated from PDRs. 
Our model calculations for PDRs were stopped where the $V$-band dust extinction, $A_\mathrm{V}$ magnitude, is five to go sufficiently in depth, following \citet{Cormier2019}.
In addition, we set a stopping criterion for the \textsc{H\,ii} region as the depth where an ionized hydrogen fraction reaches one percent \citep{Abel2005,Nagao2011}.
Note that we omit the density-bounded \textsc{H\,ii} region \citep[e.g.,][]{Beckman2000,Nakajima-Ouchi2014} in our model to decrease the number of free parameters.

We investigated three free parameters to be obtained in our model: the hydrogen gas volume density at the gas-illuminated face $n_\mathrm{H,c}$, the ionization parameter $U$ at the gas-illuminated face, and the covering fraction of a PDR $cov_\mathrm{PDR}$.
Here, $cov_\mathrm{PDR}$ is not an input parameter of \textsc{Cloudy}, and we define this as the linear combination coefficient of the fraction of the \textsc{H\,ii} region covered by a PDR.
Our modeled luminosities are given by 
\begin{equation}
    L_\mathrm{out} = \left(1-cov_\mathrm{PDR}\right)L_\mathrm{\textsc{H\,ii}}+cov_\mathrm{PDR}L_\mathrm{\textsc{H\,ii}+PDR},
	\label{eq:modelledflux}
\end{equation}
where $L_\mathrm{out}$ represents the final output luminosity, $L_\mathrm{\textsc{H\,ii}}$ corresponds to the luminosity from the \textsc{H\,ii} region completely uncovered by the PDR, and $L_\mathrm{\textsc{H\,ii}+PDR}$ denotes the luminosity from the ISM entirely covered by the PDR.
We generated a grid of models by varying the above three parameters within the range below:  $\log n_\mathrm{H,c}/\mathrm{cm^{-3}}$ in the range of $1.0\leq \log n_\mathrm{H,c}/\mathrm{cm^{-3}}\leq4.0$ at $0.1$ intervals, $\log U$ in the range of $-4.0\leq \log U\leq 0.0$ at $0.1$ intervals, and $cov_\mathrm{PDR}$ between zero (i.e., a \textsc{H\,ii} region not covered by PDR at all) and unity (i.e., a completely covered \textsc{H\,ii} region by PDR) at $0.01$ intervals. 
These parameter spaces are consistent with previous studies \citep[e.g.,][]{Cormier2019,Harikane2020}.

\subsection{Comparison with observations}

\begin{table}
\caption{Line and continuum luminosities used in our modeling.}
    \begin{center}
    \begin{tabular*}{8cm}{ll}
        \hline
        \hline
        $L_\mathrm{[\textsc{O\,iii}]88\mathrm{\mu m}}$ & $(1.29\pm0.39)\times10^9~\mathrm{L_\odot} ^\mathrm{a}$\\
        $L_\mathrm{[\textsc{C\,ii}]158\mathrm{\mu m}}$ & $(1.42\pm0.23)\times10^8~\mathrm{L_\odot} ^\mathrm{a}$\\
        $L_\mathrm{UV}$ & $(4.00\pm0.18)\times10^{10}~\mathrm{L_\odot} ^\mathrm{b}$\\
        $L_\mathrm{IR}$ & $9.8\times10^{11}~\mathrm{L_\odot} ^\mathrm{c}$ \\
        \hline
    \end{tabular*}
    \label{tab:luminosities}
    \end{center}
$^\mathrm{a}$ We calculate the line luminosities of [\textsc{O\,iii}]$~88~\mathrm{\mu m}$ and [\textsc{C\,ii}]$~158~\mathrm{\mu m}$ from the fluxes reported by \citet{Tamura2019} and \citet{Bakx2020CII}, respectively.

$^\mathrm{b}$ We calculate the UV luminosity based on the photometry of HST/F140W (rest-frame $1500$~\AA) provided by \citet{Ma2024}.

$^\mathrm{c}$ We obtain the infrared luminosity (rest-frame $8$--$1000~\mathrm{\mu m}$) based on the single dust continuum detection at $\lambda_\mathrm{rest}\approx 90~\mathrm{\mu m}$ \citep{Tamura2019} and the simple modified blackbody with the dust emissivity index $\beta_\mathrm{d}=2.0$ and a dust temperature of $80~\mathrm{K}$ from \citet{Bakx2020CII}.
We assume the error of the infrared luminosity to be $50~\%$, corresponding to a dust temperature uncertainty of $\sim\pm10~\mathrm{K}$.
\end{table}

To compare our model with the observations, we need to scale the modeled luminosities to the observed ones.
Throughout this paper, we use the magnification-corrected line and continuum luminosities, as shown in Table~\ref{tab:luminosities}.
Here, we assume the magnification factor of $1.43$ \citep{Kawamata2016, Tamura2019}.
We take the luminosity ratios instead of including the normalization factor as another free parameter.
To minimize errors involved in the ratios, we choose the \textsc{[O\,iii]}~$88~\mu\mathrm{m}$ luminosity with a better signal-to-noise ratio as the denominator.

We searched for the best-fit parameter sets by minimizing the $\chi^2$, defined as
\begin{equation}
    \chi^2=\sum^{n_\mathrm{obs}=3}_{i=1}\frac{\left(O_i - M_i\right)^2}{\sigma_i^2},
\end{equation}
where $O_i$, $M_i$, and $\sigma_i$ represent the observed and model-predicted luminosity ratios, and the error in $O_i$ for the $i$-th pair of observables, respectively.
To estimate the uncertainties of the best-fitting parameters, we employed a Monte Carlo method. 
Specifically, we repeated the fitting procedure $10,000$ times, each time perturbing the observed ratios by adding random Gaussian noise with a standard deviation equal to the measured uncertainty ($1\sigma$) of each line ratio. 
From the resulting distribution of best-fit parameter values (i.e., those minimizing $\chi^2$ in each realization), we adopted the median and $\pm34$th percentiles as the representative parameter values and their associated uncertainties.

\section{Fiducial model results} 
\label{sec:result}
\subsection{Best-fit parameters}
\label{sec:bestfitparamters}

\begin{figure}
\includegraphics[width=0.45\textwidth]{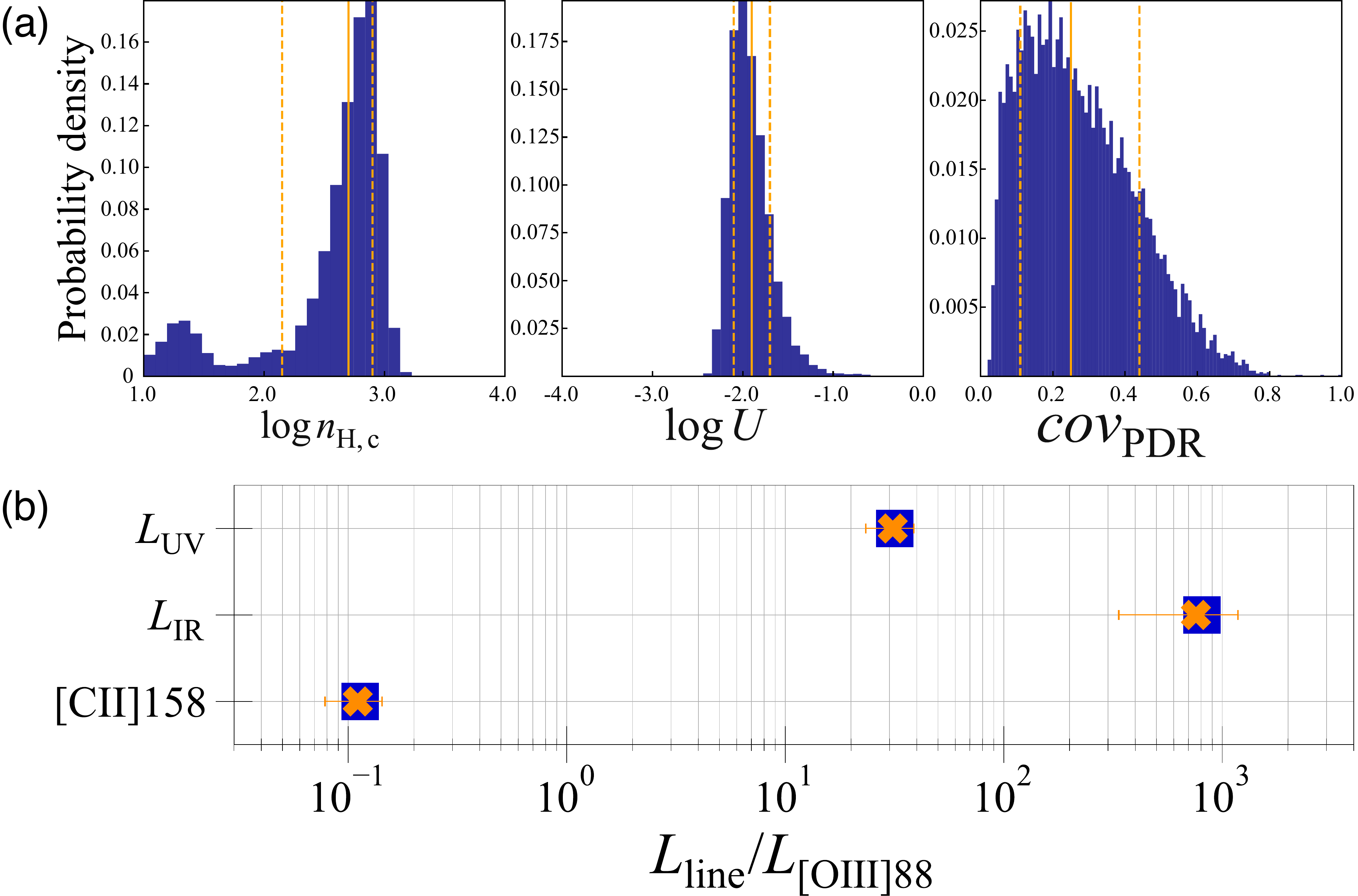}
\caption{(a) The probability density distributions of each parameter obtained from a total of $10,000$ Monte Carlo realizations. Orange solid and dashed lines correspond to the median and error as their $68$ percentiles, and we define them as the best-fitting parameters. (b) Comparison between observed (orange crosses) and modeled (blue squares) luminosity ratios relative to \textsc{[O\,iii]}$~88~\mathrm{\mu m}$.
}
\label{fig:fiducialmodel_PDF}
\end{figure}

\begin{figure}
\includegraphics[width=0.4\textwidth]{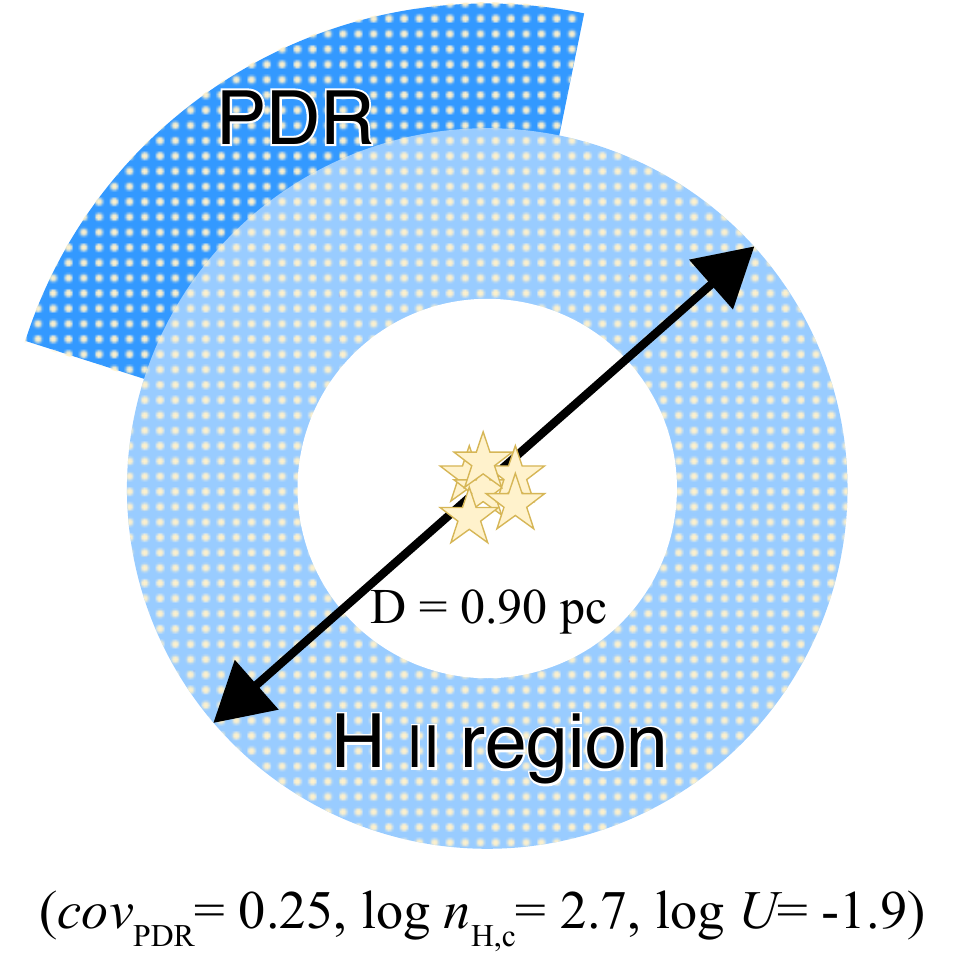}
\caption{Schematic view of the ISM for the best-fit model parameter set.
}
\label{fig:fiducialmodel_schematicview}
\end{figure}

Figure~\ref{fig:fiducialmodel_PDF} (a) shows the distributions of the best-fit parameter values obtained from each Monte Carlo realization, i.e., the parameter sets that yielded the minimum $\chi^2$ in each run. 
In the histograms, the orange solid and dashed lines indicate the median and central $68$-th percentile intervals for each parameter, respectively.
From these distributions, we derive the best-fitting parameters as $cov_\mathrm{PDR}=0.25^{+0.19}_{-0.14}$, $\log n_\mathrm{H,c}/\mathrm{cm^{-3}}=2.7^{+0.2}_{-0.55}$, and $\log U=-1.9\pm0.2$, respectively. 
In the following discussion, we use the median values of these distributions as representative parameters. 
Figure~\ref{fig:fiducialmodel_PDF} (b) compares the observed luminosity ratios with those predicted by the best-fitting model, showing good overall agreement. 
Figure~\ref{fig:fiducialmodel_schematicview} presents a two-dimensional schematic view of the `typical' ISM ionization structure derived from the best-fitting parameter set.

Our model predicts $cov_\mathrm{PDR}\approx25\%$, indicating that $\approx75\%$ of the outer surface of a typical \textsc{H\,ii} region is not covered by the PDR. 
In addition, the $cov_\mathrm{PDR}$ distribution clearly rules out a fully covered scenario ($cov_\mathrm{PDR}=1$). 
Recent studies have suggested that galaxies with a high \textsc{[O\,iii]}$~88~\mathrm{\mu m}/$\textsc{[C\,ii]}$~158~\mathrm{\mu m}$ luminosity ratio could have high Lyman continuum escape fractions \citep{Katz2022,Ura2023}. 
Non-zero escape of ionizing photons requires an ionization structure of ISM where a fraction of the \textsc{H\,ii} region is directly exposed to the intercloud space, implying a porous neutral gas structure or truncated structure with the edge of the ionization front.
Our predicted low-$cov_\mathrm{PDR}$ aligns with this picture, supporting these previous findings. 
In summary, our result implies that ISM structures enabling ionizing photons to escape into the intercloud space exist even in relatively enriched galaxies with significant dust continuum detection during the EoR.

Our model suggests a gas density of $\approx500~\mathrm{cm^{-3}}$ for the \textsc{H\,ii} region in Y1, which is consistent with the prediction by \citet{Vallini2021}.
This value is roughly two times higher than those estimated for $z\approx7$ galaxies using \textsc{[O\,iii]}$~88~\mu\mathrm{m}$ and other rest-frame FIR emission lines \citep{Sugahara2021,Killi2023}. 
Our result is in agreement with the redshift evolution over $0\lesssim z\lesssim10$ of electron density estimated from the \textsc{[O\,ii]}$\lambda\lambda3727,3729$, \textsc{[S\,ii]}$\lambda\lambda6716,6731$ doublet lines, and \textsc{[O\,iii]}$~88~\mathrm{\mu m}/$\textsc{[O\,iii]}$\lambda5008$ emission lines \citep{Abdurrouf2024}\footnote{They fit a power-law function to the relation between redshifts and electron densities, based on compiled measurements at $z=0$--$10$ using those nebular line combinations.}.
They suggested that one of the origins of the redshift evolution is the metallicity evolution of galaxies from $\sim500~\mathrm{Myr}$ after the Big Bang.
Y1 is estimated to have a metallicity of $\lesssim0.2~\mathrm{Z_\odot}$ \citep{Tamura2019, Harshan2024}, similar to those of galaxy samples ($0.1\lesssim Z/\mathrm{Z_\odot}\lesssim0.4$) at $z\gtrsim4$ used in \citet{Abdurrouf2024}.
On the other hand, the targets with FIR line-based electron density measurement are estimated to be relatively metal enriched ($Z>0.4~\mathrm{Z_\odot}$ at $z\approx7$; \citealt{Sugahara2021, Killi2023}), implying that these higher-metallicity galaxies might not follow the same relation.
Therefore, the metallicity difference between samples could explain the variations in the estimated electron densities.

%


Our estimated value for the electron density of Y1 is $\gtrsim 1.5$ times larger than that presented by \citet{Harshan2024}. 
They used the combination of the three \textsc{[O\,iii]} emission lines (rest-frame optical $4364~$\AA, $5008~$\AA\ from their observations and rest-frame FIR $88~\mathrm{\mu m}$ from \citealt{Tamura2019}) and PyNeb \citep{Luridiana2015}. 
As their NIRSpec/MSA slits covered only two of the three stellar clumps, as mentioned in Section~\ref{sec:target}, it is likely that the total intensities of the optical lines they observed are underestimated by a factor of $\approx 2/3$. 
If we simply assume that their flux measurements of optical \textsc{[O\,iii]} emission lines account for $2/3$ of the entire system, their $n_\mathrm{e}$ estimate could increase, giving an upper limit of $n_\mathrm{e}\approx500~\mathrm{cm^{-3}}$, consistent with our estimate.

We obtained an ionization parameter of $\log U\approx-2.0$, which is roughly comparable to typical ionization parameters estimated for metal-poor UV bright galaxies at $z\sim6$--$8$ from the JADES survey \citep{Cameron2023}.
In contrast, this is higher than the median value reported in the Herschel Dwarf Galaxy Survey ($-2.4$; \citealt{Cormier2019}), the average values of \textsc{H\,ii} regions in our Galaxy ($\approx-2.3$; \citealt{Rigby2004}), and observations of super-star clusters in M82 \citep{Forster-Schreier2001,Smith2006}.
The higher ionization parameter found for Y1 reflects the observed high \textsc{[O\,iii]}$~88~\mathrm{\mu m}$-to-\textsc{[C\,ii]}$~158~\mathrm{\mu m}$ luminosity ratio for Y1 compared to the local dwarf galaxies and star-forming regions. 

\citet{Ma2024} and \citet{Harshan2024} have reported two distinct values of $\log U=-1.60^{+0.14}_{-0.05}$ and $\log U=-2.48\pm0.30$, respectively.
A simplest explanation of this discrepancy is in the difference between the tracers and the method they used.
\citet{Ma2024} estimated the ionization parameter based on the SED fits of the HST/JWST photometric data with \texttt{Bagpipes}, so their result depends on their assumption for nebular conditions and stellar population models.
\citet{Harshan2024} calculated the line ratio of \textsc{[O\,iii]}$\lambda5008/$\textsc{[O\,ii]}$\lambda\lambda3727+3729$ (=$\mathrm{O32}$) and used an analytical relationship between this line ratio and $\log U$ calibrated by \citet{Kewley2019}.
Note that this relation is valid in the range of $-3.98\leq\log U\leq-2.48$, which is entirely different from the range \citet{Ma2024} explored ($-2.5<\log U<-0.5$). 
Our method involves \textsc{[C\,ii]}$~158~\mathrm{\mu m}$ and fully simulates the multi-phase ISM all the way from the ionized out to the neutral regions, which is in stark contrast to the method that \citet{Ma2024} or \citet{Harshan2024} employed.

\subsection{Predicted SED and emission line luminosities}
\label{sec:SEDprediction}

\begin{figure*}
\includegraphics[width=0.95\textwidth]{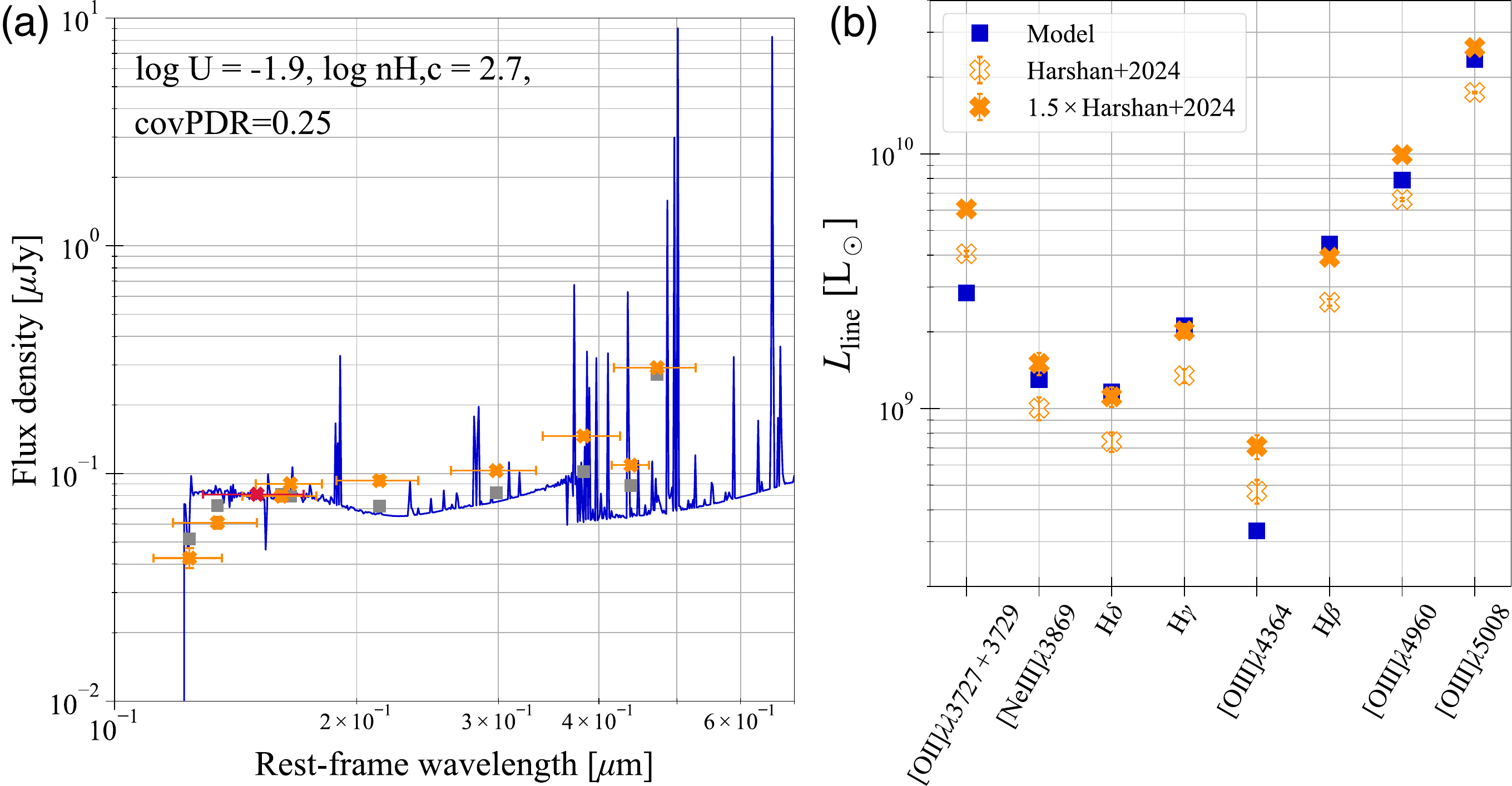}
\caption{(a) Rest-frame UV-to-Optical SED obtained from the \textsc{Cloudy} model with the best-fit parameters (blue line) with modeled flux densities (gray squares) and extracted flux densities in \citet{Ma2024} from HST and JWST observations (red and orange crosses). We used the red cross, the observed flux density of HST/F140W, to scale the model to the observations. (b) Comparison of rest-frame optical emission line luminosities between our model with the best-fit parameters (blue squares) and observations (orange crosses). We correct the reported line fluxes in \citet{Harshan2024} for dust attenuation (blank crosses). In addition, we consider luminosities with an additional factor of $1.5$, which originates from the slit covering factor in \citet{Harshan2024} (filled crosses). Our model reproduces the observed line luminosities of the Balmer lines ($\mathrm{H\beta}$, $\mathrm{H\gamma}$, $\mathrm{H\delta}$), \textsc{[O\,iii]}$\lambda\lambda4960,5008$ and [Ne\textsc{\,iii]}$\lambda3869$. However, it underestimates the luminosities of the \textsc{[O\,iii]}$\lambda4364$ and \textsc{[O\,ii]}$\lambda\lambda3727+3729$ emission lines by approximately a factor of two.
}
\label{fig:fiducialmodel_SED}
\end{figure*}

Figure~\ref{fig:fiducialmodel_SED} (a) compares the best-fitting SED predicted by \textsc{Cloudy} with the observed one, which allows for validation of our model.
The predicted SED is normalized so as to match the HST/F140W photometry reported by \citet{Ma2024}.
We find that the observed and modeled flux densities in the rest-frame FUV and JWST/F444W are in reasonable agreement.
However, our model underestimates the flux densities at $2000\lesssim\lambda_\mathrm{rest}/\mathrm{\AA}\lesssim4000$ (JWST/F200W, 277W, F356W, F410M) by $\sim 70\%$.

The possible attribution of this discrepancy is as follows.
Firstly, the dust attenuation law or grain size distribution we assumed in \textsc{Cloudy} may differ from those of Y1. 
As mentioned in Section~\ref{sec:method}, our model employs the Milky Way-like extinction law, which is known to have an enhanced extinction at $\lambda_\mathrm{rest}\approx2175~$\AA\ and could reduce the FUV flux densities for a fixed $A_\mathrm{V}$.
This absorption feature is expected in the presence of interstellar growth of carbonaceous grains, but it is not likely for Y1 as it is too young to have substantial grain growth \citep{Tamura2019}.
If this is the case, adapting a dust extinction curve without the 2175~\AA\ bump, such as the Calzetti law \citep{Calzetti2000}, could mitigate the discrepancy at F200W and F277W.
This expectation aligns with the dust extinction curve for UV-luminous LBGs predicted from the infrared-excess (IRX)-$\beta$ relation \citep{Bowler2024, Sugahara2024}.
In addition, it is in agreement with the predicted dust extinction curves for two star-forming galaxies at $z\sim6$--$8$ by \citet{Markov2023}, where they combine JWST/NIRSpec MSA data with the SED fitting technique, including parameterizations of a dust attenuation law.
The second possibility is the existence of an older stellar population in addition to the young stellar population in our \textsc{Cloudy} model, which could redden the $\lambda_\mathrm{rest} \gtrsim 4000~$\AA\ continuum spectrum.
Although \citet{Harshan2024} did not report a significant Balmer break in the NIRSpec MSA spectrum, the presence of the older population has been proposed to explain the observed high dust mass and metallicity of Y1 \citep{Tamura2019}.
We test this scenario in Section~\ref{sec:oldstellarpopulation}.

Another indicator to validate our model is a series of the optical emission lines that are well-constrained by recent JWST observations \citep{Harshan2024}.
Our best-fitting model directly predicts the intrinsic optical line luminosities.
They are scaled using the normalization factor we employed to match the model and observed SEDs (see Figure~\ref{fig:fiducialmodel_SED} (a)).
Then, they are compared with the extinction-corrected line luminosities, which were obtained from the JWST/NIRSpec MSA observations (Table~2 of \citealt{Harshan2024}) to be independent of the choice of extinction laws. 
Note that we correct the observed line luminosities by multiplying them with a factor of $1.5$ since it is likely that the NIRSpec MSA observations lost $\approx 1/3$ of the total emission, as stated in Section~\ref{sec:bestfitparamters}.
Figure~\ref{fig:fiducialmodel_SED} (b) shows the line luminosities between our model (blue squares) and the observed spectrum (orange crosses).
Without $1.5\times$ correction, our model overestimates most of (6 out of 8) optical line luminosities. 
With a $1.5\times$ correction applied, the predicted luminosities show improved agreement with six out of eight observed emission lines: $\mathrm{H\beta,\ H\gamma,\ H\delta}$, \textsc{[O\,iii]}$\lambda4960$, \textsc{[O\,iii]}$\lambda5008$, [Ne\textsc{\,iii]}$\lambda3869$.

However, our model underestimates the luminosities of the \textsc{[O\,iii]}$\lambda4364$ and \textsc{[O\,ii]}$\lambda\lambda3727+3729$ emission lines by approximately a factor of two.
The upper energy levels of the \textsc{[O\,iii]}$\lambda4364$ and \textsc{[O\,iii]}$\lambda5008$ transitions are different, and the relative excitation rates to these levels depend on the electron temperature in the low-density regime ($n_\mathrm{e}\lesssim10^{4}~\mathrm{cm^{-3}}$; \citealt{Osterbrock2006}). 
Since such densities are typical in local \textsc{H\,ii} regions \citep{Kim2001}, the line ratio between these transitions has been widely used as a diagnostic of $T_\mathrm{e}$.
\citet{Harshan2024} found an electron temperature of Y1 to be $T_\mathrm{e}=17634\pm3900~\mathrm{K}$ based on this line ratio and the best-fitting relation toward the modeled line ratio with the photoionization model obtained by \citet{Nicholles2020}.
The inferred $T_\mathrm{e}$ derived from the \textsc{[O\,iii]}$\lambda4364$/\textsc{[O\,iii]}$\lambda5008$ ratio of our best-fitting model is $T_\mathrm{e}\approx13200~\mathrm{K}$, which is lower than \citet{Harshan2024}.
We also find no model grid that reproduces such a high $T_\mathrm{e}$.

The relatively low $T_\mathrm{e}$ predicted by our model may result from the relatively high gas-phase metallicity assumed in our \textsc{Cloudy} calculations. 
The \textsc{[O\,iii]}$\lambda4364/$\textsc{[O\,iii]}$\lambda5008$ line ratio is sensitive to electron temperature, and because metals efficiently cool the gas, this ratio is negatively correlated with gas-phase metallicity \citep[see][]{Osterbrock2006, Draine2011}.
Our model assumes a metallicity of $Z=0.2~\mathrm{Z_\odot}$ or $12+\log\left(\mathrm{O/H}\right)=7.99$ if we take the solar value of $12+\log\left(\mathrm{O/H}\right)_\odot=8.69$ \citep{Grevesse2010}.
In contrast, \citet{Harshan2024} estimated an oxygen abundance of $12+\log\left(\mathrm{O/H}\right)=7.8\pm0.2$ ($Z\approx0.12~\mathrm{Z_\odot}$). 
Thus, adopting a lower metallicity in our model may mitigate the tension. 
Note that $Z\approx0.12~\mathrm{Z_\odot}$ is even consistent with our previous estimate \citep[$Z = 0.20^{+0.16}_{-0.18}~\mathrm{Z_\odot}$,][]{Tamura2019} within $1\sigma$.
On the other hand, \citet{Ma2024} reported a solar metallicity, $Z\approx Z_\odot$, for the eastern component of Y1, which was dropped in the NIRSpec MSA spectroscopy of \citet{Harshan2024}.
The flux contribution from this metal-enriched eastern component would potentially reduce the total \textsc{[O\,iii]}$\lambda4364/$\textsc{[O\,iii]}$\lambda5008$ line ratio, mitigating the discrepancy. 

The \textsc{[O\,iii]}$\lambda5008/$\textsc{[O\,ii]}$\lambda\lambda3727+3729$ emission line ratio primarily depends on the ionization parameter when the metallicity difference is relatively small \citep[e.g., a variation of $\Delta U\sim0.3~\mathrm{dex}$ between $0.1~\mathrm{Z_\odot}$ and $0.2~\mathrm{Z_\odot}$;][]{Kewley-Dopita2002}. 
Therefore, the underestimation of the \textsc{[O\,ii]}$\lambda\lambda3727+3729$ flux in our model is likely due to the high ionization parameter predicted by the best-fit solution.
However, because of the limited number of observables, our modeling assumes a single-phase ionized gas component characterized by a single set of gas density and ionization parameter. 
As a result, the current model cannot fully reproduce the observed \textsc{[O\,ii]}$\lambda\lambda3727+3729$ flux. 
To resolve this discrepancy, it may be necessary to introduce an additional ionized gas component with a lower ionization parameter (i.e., $\log U < -2.0$) beyond the framework of our fiducial model, which could account for the enhanced \textsc{[O\,ii]}$\lambda\lambda3727+3729$ emission. 
Incorporating additional observational constraints in future studies may allow for more detailed modeling of such multi-phase ionized gas structures.


\section{Disccussions}
\label{sec:discussion}

\subsection{Potential coexistence of an old stellar component}
\label{sec:oldstellarpopulation}

\begin{figure}
\includegraphics[width=0.45\textwidth]{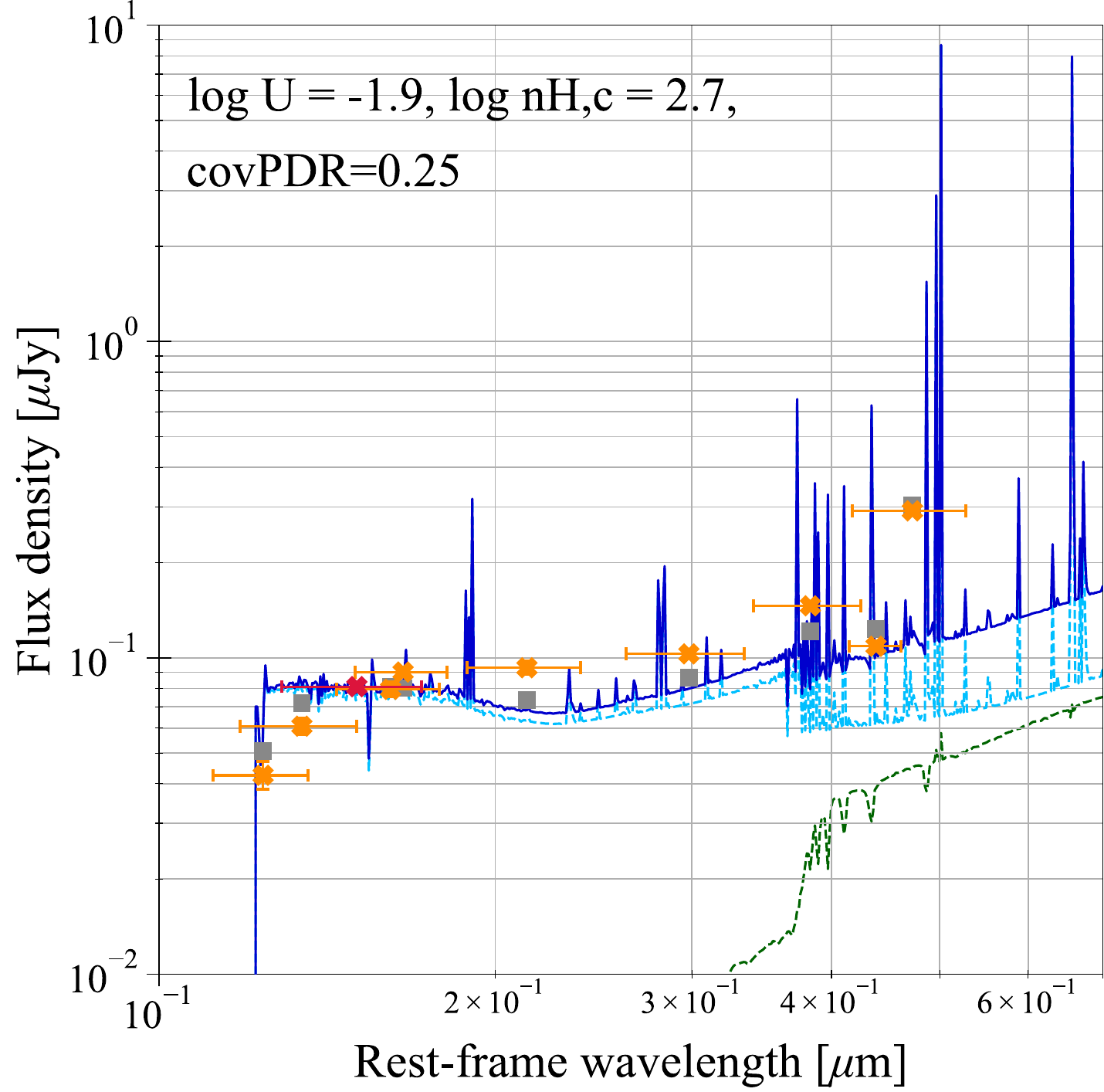}
\caption{Rest-frame UV-to-Optical SED model (blue line) in which an old stellar component made by single star-burst at $300~\mathrm{Myr}$ ago (dashed green) is added to the best-fit SED model obtained in Section~\ref{sec:SEDprediction} (dashed light-blue) with modeled flux densities (gray squares) and extracted flux densities in \citet{Ma2024} from HST and JWST observations (red and orange crosses). We used the red cross, the observed flux density of HST/F140W, to scale the model to the observations. Including the additional contribution of the old stellar component makes the difference between models and observations small in $\lambda_\mathrm{rest}\gtrsim 4000~$\AA.
}
\label{fig:additinonal_oldstarmodel_SED}
\end{figure}

One of the potential solutions for reproducing the observed photometry at $\lambda_\mathrm{rest}\gtrsim4000~\mathrm{\AA}$ is an additional contribution from the underlying stellar component, which is older than the $4~\mathrm{Myr}$-old component we used in the \textsc{Cloudy} model.
Indeed, \citet{Tamura2019} suggested a possible contribution of the stellar population with a $\sim 300~\mathrm{Myr}$ age for a necessary source of gas-phase metal and dust mass. 
In what follows, we test whether adding the old stellar population improves the discrepancy at $\lambda_\mathrm{rest}\gtrsim4000~\mathrm{\AA}$. 
In \textsc{Cloudy} calculations, we consider a stellar component that formed by a past single starburst $300$ and $600~\mathrm{Myr}$ ago at the time of $z = 8.31$, when the age of the Universe was $\approx600~\mathrm{Myr}$. 
We assume the same ionization parameter and gas density as the best-fitting parameters of the fiducial model (see Section~\ref{sec:bestfitparamters}). 
The stellar SEDs are processed with the same \textsc{Cloudy} setup as the fiducial model before being added to the young stellar component. 
The fractional contribution of an old stellar component to the young stellar component is determined by $\chi^2$ fitting to the observed photometry, as shown with the orange crosses in Figure~\ref{fig:fiducialmodel_SED} (a). 

As a result, we find that the least-$\chi^2$ values are almost the same in adding a stellar component of either $300$ or $600~\mathrm{Myr}$ ages ($\chi^2=194.34$ for $300~\mathrm{Myr}$ and $\chi^2=193.00$ for $600~\mathrm{Myr}$).
We adopt the case of adding the $300~\mathrm{Myr}$ stellar population following \citet{Tamura2019}, where the young-to-old fractional contribution is 1.7:1 in the Paschen continuum luminosity ($3646$--$8204~\mathrm{\AA}$).
Figure~\ref{fig:additinonal_oldstarmodel_SED} shows the model SEDs of young- (light blue), old- (green), and composite-stellar populations (blue) with modeled (gray squares) and observed photometry (orange and red crosses). 
We find that the F410M photometry is improved, although it still fails to reproduce F356W. 
This is likely due to the fact that the emission line contribution to F356W, such as \textsc{[O\,ii]}$\lambda\lambda3727,3729$, is significantly underestimated due to a high ionization parameter, as we saw in Section~\ref{sec:SEDprediction}.
Adding a less-ionized component is necessary for fully characterizing the ISM, which we leave for future works.

\subsection{Physical size and morphology of a typical \textsc{H\,ii} region}
\label{sec:size_HIIregion}
The spatial scale of a typical \textsc{H\,ii} region would help one understand the evolutionary phase of the star formation in Y1. 
Although resolving individual \textsc{H\,ii} regions is impossible, the nebular parameters, $n_\mathrm{H}$ and $U$, set the Str\"{o}mgren sphere with a radius of $r_\mathrm{s}$, which can be used as a proxy for a spatial scale of the typical \textsc{H\,ii} region of Y1.
We derive a hydrogen gas density of $\log n_\mathrm{H,c}/\mathrm{cm^{-3}}=2.7$, and an ionization parameter of $\log U=-1.9$ as the physical properties of the typical ISM in Y1.
In the \textsc{Cloudy} code, $r_\mathrm{s}$ is defined as the point where the neutral hydrogen fraction reaches $\mathrm{H^0}/\mathrm{H_\mathrm{tot}}=0.5$.
Here, we simply assume that the $r_\mathrm{s}$ is the radius of the modeled \textsc{H\,ii} region.
One should note that the size of the \textsc{H\,ii} region, in reality, includes the distance from the central ionizing source to the gas-illuminated face.


Our model estimates $r_\mathrm{s}=0.45~\mathrm{pc}$, corresponding to the \textsc{H\,ii} region size of $D=0.90~\mathrm{pc}$ in diameter.
This size scale is as large as a compact \textsc{H\,ii} region found in the Milky Way \citep[e.g.,][]{Kurtz2005,Urquhart2013}, suggesting that the typical \textsc{H\,ii} region in Y1 is in a relatively young evolutionary stage, consistent with its young stellar age ($\sim4~\mathrm{Myr}$).
In general, such a young \textsc{H\,ii} region does not reach pressure equilibrium, which supports our initial assumption of the continuous density profile at the ionization front. 
The derived size of $D=0.90~\mathrm{pc}$ and hydrogen density of $\log n_\mathrm{H,c}/\mathrm{cm^{-3}}=2.7$ follow the local size--density relation presented in \citet{Kim2001}.

Our model predicts two seemingly contradictory ISM properties: The porous neutral gas and the compact \textsc{H\,ii} region that is generally thought to be embedded in a molecular cloud.
\citet{Kim2001} pointed out that most local ultra-compact \textsc{H\,ii} regions are likely associated with extended emissions. 
To explain its origin, they proposed a simple geometrical model (see Figure~8 of their study) in which central ionizing sources in an ultra-compact \textsc{H\,ii} region are born at the edge of a molecular cloud and are subsequently ionizing the lower-density side faster, making the extended emission associated with the ultra-compact \textsc{H\,ii} region.
This picture is consistent with the porous geometry of neutral gas we depict in Y1 (Figure~\ref{fig:fiducialmodel_schematicview}), suggesting the presence of multiple density components, which may be characterized by a similar photoionization model constrained with more nebular lines with a range of critical densities.

\subsection{Systematics with fixed parameters}  
\label{sec:systematics}
\begin{figure}
\includegraphics[width=0.45\textwidth]{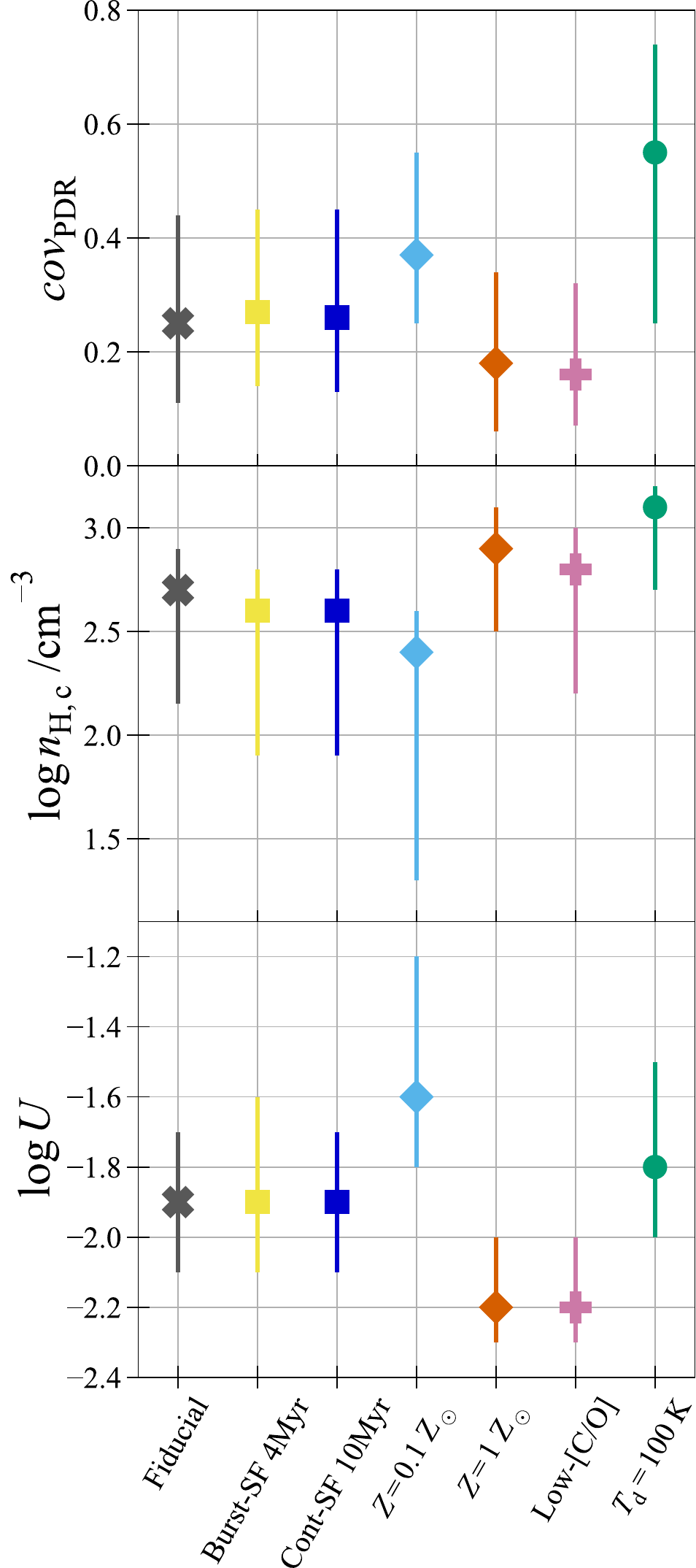}
\caption{Comparison of the best-fitting values for the covering fraction ($cov_\mathrm{PDR}$), gas density ($\log n_\mathrm{H,c}/\mathrm{cm^{-3}}$), and ionization parameter ($\log U$), with the central $68$ percentile obtained from the same method used in Section~\ref{sec:bestfitparamters}. In addition to the fiducial model, six models with different assumptions are tested (see Section~\ref{sec:systematics} for details about those models). Most of the results are consistent with the fiducial model. Although we find differences in the cases of lower-$Z$ ($Z=10\%~\mathrm{Z_\odot}$) and higher-dust temperature (i.e., higher-infrared luminosity), the differences do not change our conclusion.
}
\label{fig:resultcomparison}
\end{figure}

Due to the limited number of observables, we have to focus on only three parameters ($cov_\mathrm{PDR}$, $\log n_\mathrm{H,c}$, $\log U$) with the other parameters fixed in our modeling.
However, some of these fixed parameters could influence our conclusions obtained from the fiducial model.
In this part, we validate the reliability of our model predictions by changing parameters that are fixed in the fiducial model.
Figure~\ref{fig:resultcomparison} summarizes the best-fit value with the central $68$ percent confidence interval of each parameter predicted for a total of seven models, including the fiducial one. 
Below, we explain the details of the model parameters and the fitting results. 

\begin{figure}
\includegraphics[width=0.45\textwidth]{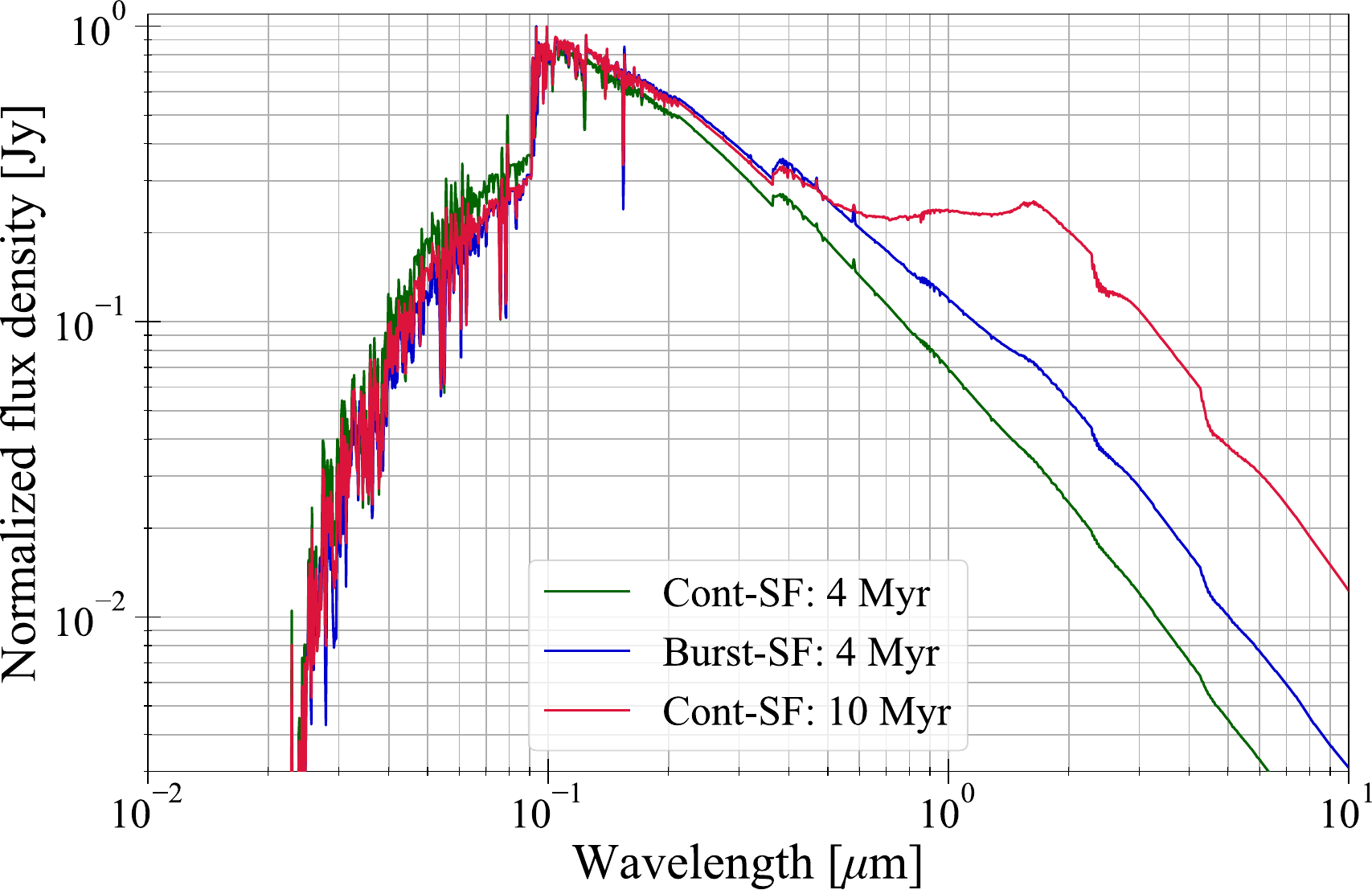}
\caption{Comparison of normalized rest-frame UV-to-MIR SED of input stellar models. We find no significant difference between continuous and bursty SFH models of $4~\mathrm{Myr}$ and also between the stellar age of $4~\mathrm{Myr}$ and $10~\mathrm{Myr}$. We assumed $\log n_\mathrm{H,c}/\mathrm{cm^{-3}}=3.0$ and $\log U=-2.0$ for all SEDs.
}
\label{fig:stellarSEDcomparison}
\end{figure}

\paragraph{SFH and stellar age}
Our fiducial model adopts a continuous star-formation of $4~\mathrm{Myr}$ with the BPASS code.
Here, we check the effects on our model predictions by changing (i) star-formation history (SFH) to bursty one and (ii) the stellar age to $10~\mathrm{Myr}$ following \citet{Inoue2014} and \citet{Cormier2019}. 
Figure~\ref{fig:stellarSEDcomparison} compares the rest-frame UV-to-MIR input stellar SED among them, normalized to unity at the peak for all models. 
We find no significant difference in the spectral shape of the UV range among these three models, suggesting no significant change in our model results.
The yellow and blue squares in Figure~\ref{fig:resultcomparison} correspond to the best-fit parameters for the bursty-SFH of $4~\mathrm{Myr}$ and constant SFH of $10~\mathrm{Myr}$, respectively.
We find that the obtained best-fit parameters are consistent with each other; therefore, the differences in SFH and stellar ages do not change our model predictions.

\paragraph{Metallicity dependence}
We assumed that the gas and stellar metallicities of Y1 are the same, with $Z=0.2~\mathrm{Z_\odot}$, based on the SED fit results shown by \citet{Tamura2019} in our fiducial model.
However, differences in metallicity generally impact line diagnostics \citep[e.g.,][]{Osterbrock2006, Nagao2011}.
Here, we test two cases: $Z=0.10~\mathrm{Z_\odot}$, which is consistent with the prediction by \citet{Harshan2024} based on optical emission lines, and $Z=1.0~\mathrm{Z_\odot}$, used as the highest metallicity case for galaxies at $z\gtrsim6$ in previous studies \citep[e.g.,][]{Harikane2020, Sugahara2022}.
The light blue and orange diamonds in Figure~\ref{fig:resultcomparison} represent the best-fit parameters with the central $68$ percent confidence intervals for $Z=0.10~\mathrm{Z_\odot}$ and $Z=1.0~\mathrm{Z_\odot}$, respectively.
In both cases, the predicted values for all parameters are consistent with the fiducial model within $1\sigma$.
However, we observe a trend indicating a shift in the estimated parameter ranges.
For $Z=1.0~\mathrm{Z_\odot}$, the best-fitting values of the ionization parameter and $cov_\mathrm{PDR}$ become lower, while the gas density is higher than in the fiducial model. 
Conversely, a higher $cov_\mathrm{PDR}$, a higher ionization parameter, and a lower gas density are preferred in the case of $Z=0.1~\mathrm{Z_\odot}$.
A lower $Z$ leads to a lower luminosity of \textsc{[O\,iii]}$~88~\mathrm{\mu m}$ \citep{Harikane2020, Fujimoto2024}, prompting the best-fitting values to compensate for it: A lower $\log n_\mathrm{H,c}$ increases the luminosity of \textsc{[O\,iii]}$~88~\mathrm{\mu m}$ as the gas density approaches or falls below the critical density of this line ($510~\mathrm{cm^{-3}}$ at $T_\mathrm{e}=10000~\mathrm{K}$, \citealt{Osterbrock2006}), and a higher $\log U$ increases the $L_{\textsc{[O\,iii]}88\mathrm{\mu m}}/L_{\textsc{[C\,ii]}158\mathrm{\mu m}}$ ratio \citep[see also][]{Ura2023}, and vice versa.
A higher $\log U$ also leads to a higher UV luminosity, resulting in a slightly higher $cov_\mathrm{PDR}$ to keep $L_\mathrm{UV}/L_\mathrm{[OIII]88}$ constant by absorbing more UV continuum in the lower $Z$ case and vice versa.
This test again confirms the importance of metallicity in characterizing the other nebular parameters.
We should note that the estimated gas density for the $Z=0.10~\mathrm{Z_\odot}$ model agrees with the electron density obtained by \citet{Harshan2024}, but the estimated ionization parameter is higher than theirs. 
This result supports that Y1 may have additional low-ionized gas components other than our modeled one (see Section~\ref{sec:SEDprediction}).

\paragraph{Carbon-to-Oxygen abundance ratio}
To explain high-\textsc{[O\,iii]}/\textsc{[C\,ii]}, the Carbon-to-Oxygen abundance ratio ($\mathrm{[C/O]}$) is another critical indicator \citep{Katz2022,Nyhagen2024}.
Here, we estimate the physical properties for a lower $\mathrm{[C/O]}$ ratio than the solar value with a metallicity of $Z=0.2~\mathrm{Z_\odot}$ using the scaling relation in \citet{Dopita2006},
\begin{equation}
    \mathrm{[C/H]}=6.0\times10^{-5}(Z/\mathrm{Z_\odot})+2.0\times10^{-4}(Z/\mathrm{Z_\odot})^2.
\end{equation}
The obtained $\mathrm{[C/O]}$ is $\approx37\%$ of the solar value, consistent with the estimated value for a galaxy at $z=6.23$ based on JWST/NIRSpec measurements \citep{Jones2023}.
The pink plus sign in Figure~\ref{fig:resultcomparison} shows the modeling result. 
Although the $cov_\mathrm{PDR}$ and $\log U$ are slightly lower than the fiducial model to recover the reduced \textsc{[C\,ii]}$~158~\mathrm{\mu m}$ flux, they are consistent with each other.
Therefore, we find that the difference in $\mathrm{[C/O]}$ does not change our model results.

\paragraph{Dust temperature uncertainty}
Tightly constraining the dust temperatures of high-$z$ galaxies is generally difficult due to the limited number of rest-frame FIR continuum measurement points \citep[e.g.,][]{Inami2022,Witstok2022}.
So far, this has only been achieved in two cases, A1689-zD1 at $z=7.13$ \citep{Bakx2021,Akins2022} and REBELS-25 at $z=7.31$ \citep{Algera2024REBELS25}.  
Uncertainty in the dust temperature leads to uncertainty in infrared luminosity.
For Y1, several studies consistently show a high $T_\mathrm{d}$ of $T_\mathrm{d}\gtrsim80~\mathrm{K}$ \citep{Bakx2020CII,Sommovigo2022,Fudamoto2023,Algera2024} with various methods.
As the infrared luminosity increases with $T_\mathrm{d}$ given a fixed dust emissivity index $\beta_\mathrm{d}$, it is essential to check the impact of a higher $T_\mathrm{d}$ than the fiducial model of $T_\mathrm{d}=80~\mathrm{K}$ on our results.
Here, we investigate the case of $T_\mathrm{d}=100~\mathrm{K}$, which corresponds to $L_\mathrm{IR}=2.3\times10^{12}~\mathrm{L_\odot}$.
The green circle in Figure~\ref{fig:resultcomparison} shows the best-fit parameters for this scenario.
Although the obtained $cov_\mathrm{PDR}=0.55$ is consistent with the fiducial model due to the large error, it is the largest value in explored space.
In addition, the ionization parameter does not change, unlike in the case of low-metallicity.
This result highlights the importance of dust continuum observations toward galaxies at the EoR in the (sub)mm regime to constrain the neutral gas covering fraction.
Even in this case, the solution of $cov_\mathrm{PDR}=1$ is rejected, meaning our conclusion that a part of the \textsc{H\,ii} region is exposed to the intercloud space remains unchanged.

\section{Summary}
\label{sec:summary}
We have investigated the physical properties of the typical ISM of MACS0416\_Y1 at $z=8.312$ by combining ALMA, HST, and JWST observations with photoionization models, taking into account the neutral gas porosity.
The median with error represented by $68$ percentiles is $cov_\mathrm{PDR}=0.25^{+0.19}_{-0.14}$, $\log n_\mathrm{H,c}/\mathrm{cm^{-3}}=2.7^{+0.2}_{-0.55}$, and $\log U=-1.9\pm0.2$, respectively. 
$cov_\mathrm{PDR}$ of $\approx0.25$ means that $\sim75\%$ of the outer surface of a typical \textsc{H\,ii} region is exposed to the intercloud space.
This porous neutral gas structure meets one of the requirements for non-zero escape fractions, suggesting that galaxies similar to Y1, with high-\textsc{[O\,iii]}/\textsc{[C\,ii]} ratios, could contribute to the cosmic reionization. 

We predict the rest-frame optical SED and line luminosities of Y1 based on the best-fitting parameter sets and compare them with recent observations by JWST. 
As a result, we could roughly reproduce the observed SED.
However, our model underestimates the observed photometry at $2000\lesssim\lambda_\mathrm{rest}/\mathrm{\r{A}}\lesssim4000$ and luminosities of \textsc{[O\,iii]}$\lambda4364$ and \textsc{[O\,ii]}$\lambda\lambda3727+3729$ emission lines by half. 
The discrepancies in F200W and F277W filters can be explained by the significant bump at $\lambda_\mathrm{rest}\approx2175~\mathrm{\AA}$ in the assumed dust extinction law of the MW. 
The discrepancy in F410M can be explained by additional contributions of a 300-Myr-old stellar population.
However, our model remains to underestimate the photometry of F356W even in adding such old stellar populations, indicating \textsc{[O\,ii]}$\lambda\lambda3727+3729$ emission lines significantly impact that photometry.
Therefore, we need to consider an additional ionized gas component with a low-ionization parameter to explain them consistently.

We characterize the size of the typical \textsc{H\,ii} region in Y1 with the best-fit parameters by assuming that the size $D$ of the \textsc{H\,ii} region equals $D=2r_\mathrm{s}$, where $r_\mathrm{s}$ is the Str\"{o}mgren radius.
We find $D=0.90~\mathrm{pc}$, corresponding to the size of local compact \textsc{H\,ii} regions.
The result indicates that the typical \textsc{H\,ii} region in Y1 is in a relatively young evolutionary phase.
This finding, along with the low-$cov_\mathrm{PDR}$, suggests that the central ionizing sources in \textsc{H\,ii} regions were born at the edge of the molecular clump and have begun ionizing the low-density side.

In the fiducial model, we fixed physical parameters other than $cov_\mathrm{PDR}$, $\log n_\mathrm{H,c}$, and $\log U$ due to the limited number of observables, which are the \textsc{[O\,iii]}~$88~\mu\mathrm{m}$, the \textsc{[C\,ii]}~$158~\mu\mathrm{m}$, the UV, and the infrared luminosities.
As a result of our tests in cases where these fixed parameters are changed, we found that different metallicity ($Z=0.10~\mathrm{Z_\odot}$ and $Z=1.0~\mathrm{Z_\odot}$) and higher-dust temperature ($T_\mathrm{d}=100~\mathrm{K}$) cases impact our model results.
In particular, higher $T_\mathrm{d}$ likely provides a significantly higher $cov_\mathrm{PDR}$ solution, indicating that dust continuum observations by (sub)mm facilities are crucial for constraining the neutral gas porosity.
However, even in this case, our conclusion that the typical ISM in Y1 has a porous neutral gas structure remains unchanged.

This kind of research has been limited to the local Universe. 
However, ALMA has opened a new window to high-$z$ galaxies with redshifted rest-frame FIR emission line observations. 
These rich data allow us to extend this research to other galaxies and investigate the $cov_\mathrm{PDR}$ for EoR.
In addition, JWST has provided us with data on the rest-frame UV-to-optical regimes for $z>6$ galaxies.
By taking advantage of joint observations from ALMA and JWST, we can construct more complex models, such as multi-component \textsc{H\,ii} regions.
In the upcoming years, we aim to start rest-frame FIR emission line observations toward $z>8$ galaxies in the northern hemisphere using a new wideband (sub)mm receiver, FINER \citep{Tamura2024FINER}, that will be mounted on the Large Millimeter Telescope (LMT: \citealt{Hughes2020}) in M\'{e}xico.
This will offer further opportunities to combine research with rest-frame UV-to-optical and FIR observations.

\section*{Acknowledgments}
We thank anonymous referee for the helpful comments.
M.H. was supported by the Japan Society for the Promotion of Science (JSPS) KAKENHI Grant Nos.\ 22J21948, 22KJ1598 and the ALMA Japan Research Grant of NAOJ ALMA Project, NAOJ-ALMA-360.
Y.T. and T.B. were supported by NAOJ ALMA Scientific Research Grant No.\ 2018-09B.
Y.T. is supported by JSPS KAKENHI Grant Nos.\ 17H06130, 22H04939.
H.U. acknowledges support from JSPS KAKENHI Grant Nos.\ 20H01953, 22KK0231, 23K20240, 25K01039.
T.H. was supported by JSPS KAKENHI Grant Nos.\ 22H01258 and 23K22529. 
K.M. acknowledges financial support from the JSPS through KAKENHI grant No.\ 20K14516. 
K.M. and A.K.I are supported by JSPS KAKENHI Grant No.\ 23H00131. 
A.K.I and Y.S. are supported by NAOJ ALMA Scientific Research Grant No. 2020-16B.

This paper makes use of the following ALMA data: ADS/JAO.ALMA \#2016.1.00117.S, ADS/JAO.ALMA \#2017.1.00225.S, ADS/JAO.ALMA \#2017.1.00486.S, ADS/JAO.ALMA \#2018.1.01241.S. ALMA is a partnership of ESO (representing its member states), NSF (USA) and NINS (Japan), together with NRC (Canada), MOST and ASIAA (Taiwan), and KASI (Republic of Korea), in cooperation with the Republic of Chile. The Joint ALMA Observatory is operated by ESO, AUI/NRAO and NAOJ. In addition, publications from NA authors must include the standard NRAO acknowledgement: The National Radio Astronomy Observatory is a facility of the National Science Foundation operated under cooperative agreement by Associated Universities, Inc.
HST data presented in this paper was obtained from the Mikulski Archive for Space Telescopes (MAST) at the Space Telescope Science Institute (STScI). 
The specific observations analyzed can be accessed via \dataset[10.17909/T9KK5N]{https://doi.org/10.17909/T9KK5N}.
STScI is operated by the Association of Universities for Research in Astronomy, Inc., under NASA contract NAS5-26555.
Data analysis was in part carried out on the Multi-wavelength Data Analysis System operated by the Astronomy Data Center (ADC), National Astronomical Observatory of Japan.


%

\vspace{5mm}
\facilities{ALMA, HST (WFC3).}


\software{Cloudy \citep{Ferland2017}, 
          Astropy \citep{astropy2013, astropy2018, Astropy2022},
          matplotlib \citep{Hunter07},
          NumPy \citep{Harris20},
          pandas \citep{mckinney-proc-scipy-2010},
          Scipy \citep{Virtanen2020},
          xarray \citep{Hoyer17}.
          }

\bibliography{reference}{}

\begin{thebibliography}{}
\expandafter\ifx\csname natexlab\endcsname\relax\def\natexlab#1{#1}\fi
\providecommand{\url}[1]{\href{#1}{#1}}
\providecommand{\dodoi}[1]{doi:~\href{http://doi.org/#1}{\nolinkurl{#1}}}
\providecommand{\doeprint}[1]{\href{http://ascl.net/#1}{\nolinkurl{http://ascl.net/#1}}}
\providecommand{\doarXiv}[1]{\href{https://arxiv.org/abs/#1}{\nolinkurl{https://arxiv.org/abs/#1}}}

\bibitem[{{Abdurro'uf} {et~al.}(2024){Abdurro'uf}, {Larson}, {Coe}, {Hsiao},
  {{\'A}lvarez-M{\'a}rquez}, {Crespo G{\'o}mez}, {Adamo}, {Bhatawdekar}, {Bik},
  {Bradley}, {Conselice}, {Dayal}, {Diego}, {Fujimoto}, {Furtak}, {Hutchison},
  {Jung}, {Killi}, {Kokorev}, {Mingozzi}, {Norman}, {Resseguier}, {Ricotti},
  {Rigby}, {Vanzella}, {Welch}, {Windhorst}, {Xu}, \& {Zitrin}}]{Abdurrouf2024}
{Abdurro'uf}, {Larson}, R.~L., {Coe}, D., {et~al.} 2024, \apj, 973, 47,
  \dodoi{10.3847/1538-4357/ad6001}

\bibitem[{{Abel} {et~al.}(2005){Abel}, {Ferland}, {Shaw}, \& {van
  Hoof}}]{Abel2005}
{Abel}, N.~P., {Ferland}, G.~J., {Shaw}, G., \& {van Hoof}, P.~A.~M. 2005,
  \apjs, 161, 65, \dodoi{10.1086/432913}

\bibitem[{{Akins} {et~al.}(2022){Akins}, {Fujimoto}, {Finlator}, {Watson},
  {Knudsen}, {Richard}, {Bakx}, {Hashimoto}, {Inoue}, {Matsuo},
  {Micha{\l}owski}, \& {Tamura}}]{Akins2022}
{Akins}, H.~B., {Fujimoto}, S., {Finlator}, K., {et~al.} 2022, \apj, 934, 64,
  \dodoi{10.3847/1538-4357/ac795b}

\bibitem[{{Algera} {et~al.}(2023){Algera}, {Inami}, {Oesch}, {Sommovigo},
  {Bouwens}, {Topping}, {Schouws}, {Stefanon}, {Stark}, {Aravena}, {Barrufet},
  {da Cunha}, {Dayal}, {Endsley}, {Ferrara}, {Fudamoto}, {Gonzalez},
  {Graziani}, {Hodge}, {Hygate}, {de Looze}, {Nanayakkara}, {Schneider}, \&
  {van der Werf}}]{Algera2023}
{Algera}, H. S.~B., {Inami}, H., {Oesch}, P.~A., {et~al.} 2023, \mnras, 518,
  6142, \dodoi{10.1093/mnras/stac3195}

\bibitem[{{Algera} {et~al.}(2024{\natexlab{a}}){Algera}, {Inami}, {Sommovigo},
  {Fudamoto}, {Schneider}, {Graziani}, {Dayal}, {Bouwens}, {Aravena}, {da
  Cunha}, {Ferrara}, {Hygate}, {van Leeuwen}, {De Looze}, {Palla},
  {Pallottini}, {Smit}, {Stefanon}, {Topping}, \& {van der Werf}}]{Algera2024}
{Algera}, H. S.~B., {Inami}, H., {Sommovigo}, L., {et~al.} 2024{\natexlab{a}},
  \mnras, 527, 6867, \dodoi{10.1093/mnras/stad3111}

\bibitem[{{Algera} {et~al.}(2024{\natexlab{b}}){Algera}, {Inami}, {De Looze},
  {Ferrara}, {Hirashita}, {Aravena}, {Bakx}, {Bouwens}, {Bowler}, {Da Cunha},
  {Dayal}, {Fudamoto}, {Hodge}, {Hygate}, {van Leeuwen}, {Nanayakkara},
  {Palla}, {Pallottini}, {Rowland}, {Smit}, {Sommovigo}, {Stefanon}, {Vijayan},
  \& {van der Werf}}]{Algera2024REBELS25}
{Algera}, H. S.~B., {Inami}, H., {De Looze}, I., {et~al.} 2024{\natexlab{b}},
  \mnras, 533, 3098, \dodoi{10.1093/mnras/stae1994}

\bibitem[{{Arata} {et~al.}(2020){Arata}, {Yajima}, {Nagamine}, {Abe}, \&
  {Khochfar}}]{Arata2020}
{Arata}, S., {Yajima}, H., {Nagamine}, K., {Abe}, M., \& {Khochfar}, S. 2020,
  \mnras, 498, 5541, \dodoi{10.1093/mnras/staa2809}

\bibitem[{{Astropy Collaboration} {et~al.}(2013){Astropy Collaboration},
  {Robitaille}, {Tollerud}, {Greenfield}, {Droettboom}, {Bray}, {Aldcroft},
  {Davis}, {Ginsburg}, {Price-Whelan}, {Kerzendorf}, {Conley}, {Crighton},
  {Barbary}, {Muna}, {Ferguson}, {Grollier}, {Parikh}, {Nair}, {Unther},
  {Deil}, {Woillez}, {Conseil}, {Kramer}, {Turner}, {Singer}, {Fox}, {Weaver},
  {Zabalza}, {Edwards}, {Azalee Bostroem}, {Burke}, {Casey}, {Crawford},
  {Dencheva}, {Ely}, {Jenness}, {Labrie}, {Lim}, {Pierfederici}, {Pontzen},
  {Ptak}, {Refsdal}, {Servillat}, \& {Streicher}}]{astropy2013}
{Astropy Collaboration}, {Robitaille}, T.~P., {Tollerud}, E.~J., {et~al.} 2013,
  \aap, 558, A33, \dodoi{10.1051/0004-6361/201322068}

\bibitem[{{Astropy Collaboration} {et~al.}(2018){Astropy Collaboration},
  {Price-Whelan}, {Sip{\H{o}}cz}, {G{\"u}nther}, {Lim}, {Crawford}, {Conseil},
  {Shupe}, {Craig}, {Dencheva}, {Ginsburg}, {VanderPlas}, {Bradley},
  {P{\'e}rez-Su{\'a}rez}, {de Val-Borro}, {Aldcroft}, {Cruz}, {Robitaille},
  {Tollerud}, {Ardelean}, {Babej}, {Bach}, {Bachetti}, {Bakanov}, {Bamford},
  {Barentsen}, {Barmby}, {Baumbach}, {Berry}, {Biscani}, {Boquien}, {Bostroem},
  {Bouma}, {Brammer}, {Bray}, {Breytenbach}, {Buddelmeijer}, {Burke},
  {Calderone}, {Cano Rodr{\'\i}guez}, {Cara}, {Cardoso}, {Cheedella}, {Copin},
  {Corrales}, {Crichton}, {D'Avella}, {Deil}, {Depagne}, {Dietrich}, {Donath},
  {Droettboom}, {Earl}, {Erben}, {Fabbro}, {Ferreira}, {Finethy}, {Fox},
  {Garrison}, {Gibbons}, {Goldstein}, {Gommers}, {Greco}, {Greenfield},
  {Groener}, {Grollier}, {Hagen}, {Hirst}, {Homeier}, {Horton}, {Hosseinzadeh},
  {Hu}, {Hunkeler}, {Ivezi{\'c}}, {Jain}, {Jenness}, {Kanarek}, {Kendrew},
  {Kern}, {Kerzendorf}, {Khvalko}, {King}, {Kirkby}, {Kulkarni}, {Kumar},
  {Lee}, {Lenz}, {Littlefair}, {Ma}, {Macleod}, {Mastropietro}, {McCully},
  {Montagnac}, {Morris}, {Mueller}, {Mumford}, {Muna}, {Murphy}, {Nelson},
  {Nguyen}, {Ninan}, {N{\"o}the}, {Ogaz}, {Oh}, {Parejko}, {Parley}, {Pascual},
  {Patil}, {Patil}, {Plunkett}, {Prochaska}, {Rastogi}, {Reddy Janga},
  {Sabater}, {Sakurikar}, {Seifert}, {Sherbert}, {Sherwood-Taylor}, {Shih},
  {Sick}, {Silbiger}, {Singanamalla}, {Singer}, {Sladen}, {Sooley},
  {Sornarajah}, {Streicher}, {Teuben}, {Thomas}, {Tremblay}, {Turner},
  {Terr{\'o}n}, {van Kerkwijk}, {de la Vega}, {Watkins}, {Weaver}, {Whitmore},
  {Woillez}, {Zabalza}, \& {Astropy Contributors}}]{astropy2018}
{Astropy Collaboration}, {Price-Whelan}, A.~M., {Sip{\H{o}}cz}, B.~M., {et~al.}
  2018, \aj, 156, 123, \dodoi{10.3847/1538-3881/aabc4f}

\bibitem[{{Astropy Collaboration} {et~al.}(2022){Astropy Collaboration},
  {Price-Whelan}, {Lim}, {Earl}, {Starkman}, {Bradley}, {Shupe}, {Patil},
  {Corrales}, {Brasseur}, {N{\"o}the}, {Donath}, {Tollerud}, {Morris},
  {Ginsburg}, {Vaher}, {Weaver}, {Tocknell}, {Jamieson}, {van Kerkwijk},
  {Robitaille}, {Merry}, {Bachetti}, {G{\"u}nther}, {Aldcroft},
  {Alvarado-Montes}, {Archibald}, {B{\'o}di}, {Bapat}, {Barentsen},
  {Baz{\'a}n}, {Biswas}, {Boquien}, {Burke}, {Cara}, {Cara}, {Conroy},
  {Conseil}, {Craig}, {Cross}, {Cruz}, {D'Eugenio}, {Dencheva}, {Devillepoix},
  {Dietrich}, {Eigenbrot}, {Erben}, {Ferreira}, {Foreman-Mackey}, {Fox},
  {Freij}, {Garg}, {Geda}, {Glattly}, {Gondhalekar}, {Gordon}, {Grant},
  {Greenfield}, {Groener}, {Guest}, {Gurovich}, {Handberg}, {Hart},
  {Hatfield-Dodds}, {Homeier}, {Hosseinzadeh}, {Jenness}, {Jones}, {Joseph},
  {Kalmbach}, {Karamehmetoglu}, {Ka{\l}uszy{\'n}ski}, {Kelley}, {Kern},
  {Kerzendorf}, {Koch}, {Kulumani}, {Lee}, {Ly}, {Ma}, {MacBride}, {Maljaars},
  {Muna}, {Murphy}, {Norman}, {O'Steen}, {Oman}, {Pacifici}, {Pascual},
  {Pascual-Granado}, {Patil}, {Perren}, {Pickering}, {Rastogi}, {Roulston},
  {Ryan}, {Rykoff}, {Sabater}, {Sakurikar}, {Salgado}, {Sanghi}, {Saunders},
  {Savchenko}, {Schwardt}, {Seifert-Eckert}, {Shih}, {Jain}, {Shukla}, {Sick},
  {Simpson}, {Singanamalla}, {Singer}, {Singhal}, {Sinha}, {Sip{\H{o}}cz},
  {Spitler}, {Stansby}, {Streicher}, {{\v{S}}umak}, {Swinbank}, {Taranu},
  {Tewary}, {Tremblay}, {de Val-Borro}, {Van Kooten}, {Vasovi{\'c}}, {Verma},
  {de Miranda Cardoso}, {Williams}, {Wilson}, {Winkel}, {Wood-Vasey}, {Xue},
  {Yoachim}, {Zhang}, {Zonca}, \& {Astropy Project Contributors}}]{Astropy2022}
{Astropy Collaboration}, {Price-Whelan}, A.~M., {Lim}, P.~L., {et~al.} 2022,
  \apj, 935, 167, \dodoi{10.3847/1538-4357/ac7c74}

\bibitem[{{Bakx} {et~al.}(2020){Bakx}, {Tamura}, {Hashimoto}, {Inoue}, {Lee},
  {Mawatari}, {Ota}, {Umehata}, {Zackrisson}, {Hatsukade}, {Kohno}, {Matsuda},
  {Matsuo}, {Okamoto}, {Shibuya}, {Shimizu}, {Taniguchi}, \&
  {Yoshida}}]{Bakx2020CII}
{Bakx}, T. J.~L.~C., {Tamura}, Y., {Hashimoto}, T., {et~al.} 2020, \mnras, 493,
  4294, \dodoi{10.1093/mnras/staa509}

\bibitem[{{Bakx} {et~al.}(2021){Bakx}, {Sommovigo}, {Carniani}, {Ferrara},
  {Akins}, {Fujimoto}, {Hagimoto}, {Knudsen}, {Pallottini}, {Tamura}, \&
  {Watson}}]{Bakx2021}
{Bakx}, T. J.~L.~C., {Sommovigo}, L., {Carniani}, S., {et~al.} 2021, \mnras,
  508, L58, \dodoi{10.1093/mnrasl/slab104}

\bibitem[{{Bakx} {et~al.}(2024){Bakx}, {Algera}, {Venemans}, {Sommovigo},
  {Fujimoto}, {Carniani}, {Hagimoto}, {Hashimoto}, {Inoue}, {Salak},
  {Serjeant}, {Vallini}, {Eales}, {Ferrara}, {Fudamoto}, {Imamura}, {Inoue},
  {Knudsen}, {Matsuo}, {Sugahara}, {Tamura}, {Taniguchi}, \&
  {Yamanaka}}]{Bakx2024}
{Bakx}, T. J.~L.~C., {Algera}, H. S.~B., {Venemans}, B., {et~al.} 2024, \mnras,
  532, 2270, \dodoi{10.1093/mnras/stae1613}

\bibitem[{{Beckman} {et~al.}(2000){Beckman}, {Rozas}, {Zurita}, {Watson}, \&
  {Knapen}}]{Beckman2000}
{Beckman}, J.~E., {Rozas}, M., {Zurita}, A., {Watson}, R.~A., \& {Knapen},
  J.~H. 2000, \aj, 119, 2728, \dodoi{10.1086/301380}

\bibitem[{{Behroozi} {et~al.}(2019){Behroozi}, {Wechsler}, {Hearin}, \&
  {Conroy}}]{Behroozi2019}
{Behroozi}, P., {Wechsler}, R.~H., {Hearin}, A.~P., \& {Conroy}, C. 2019,
  \mnras, 488, 3143, \dodoi{10.1093/mnras/stz1182}

\bibitem[{{Bouwens} {et~al.}(2022){Bouwens}, {Smit}, {Schouws}, {Stefanon},
  {Bowler}, {Endsley}, {Gonzalez}, {Inami}, {Stark}, {Oesch}, {Hodge},
  {Aravena}, {da Cunha}, {Dayal}, {de Looze}, {Ferrara}, {Fudamoto},
  {Graziani}, {Li}, {Nanayakkara}, {Pallottini}, {Schneider}, {Sommovigo},
  {Topping}, {van der Werf}, {Algera}, {Barrufet}, {Hygate}, {Labb{\'e}},
  {Riechers}, \& {Witstok}}]{Bouwens2022}
{Bouwens}, R.~J., {Smit}, R., {Schouws}, S., {et~al.} 2022, \apj, 931, 160,
  \dodoi{10.3847/1538-4357/ac5a4a}

\bibitem[{{Bowler} {et~al.}(2024){Bowler}, {Inami}, {Sommovigo}, {Smit},
  {Algera}, {Aravena}, {Barrufet}, {Bouwens}, {da Cunha}, {Cullen}, {Dayal},
  {De Looze}, {Dunlop}, {Fudamoto}, {Mauerhofer}, {McLure}, {Stefanon},
  {Schneider}, {Ferrara}, {Graziani}, {Hodge}, {Nanayakkara}, {Palla},
  {Schouws}, {Stark}, \& {van der Werf}}]{Bowler2024}
{Bowler}, R.~A.~A., {Inami}, H., {Sommovigo}, L., {et~al.} 2024, \mnras, 527,
  5808, \dodoi{10.1093/mnras/stad3578}

\bibitem[{{Bunker} {et~al.}(2023){Bunker}, {Saxena}, {Cameron}, {Willott},
  {Curtis-Lake}, {Jakobsen}, {Carniani}, {Smit}, {Maiolino}, {Witstok},
  {Curti}, {D'Eugenio}, {Jones}, {Ferruit}, {Arribas}, {Charlot}, {Chevallard},
  {Giardino}, {de Graaff}, {Looser}, {L{\"u}tzgendorf}, {Maseda}, {Rawle},
  {Rix}, {Del Pino}, {Alberts}, {Egami}, {Eisenstein}, {Endsley}, {Hainline},
  {Hausen}, {Johnson}, {Rieke}, {Rieke}, {Robertson}, {Shivaei}, {Stark},
  {Sun}, {Tacchella}, {Tang}, {Williams}, {Willmer}, {Baker}, {Baum},
  {Bhatawdekar}, {Bowler}, {Boyett}, {Chen}, {Circosta}, {Helton}, {Ji},
  {Kumari}, {Lyu}, {Nelson}, {Parlanti}, {Perna}, {Sandles}, {Scholtz},
  {Suess}, {Topping}, {{\"U}bler}, {Wallace}, \& {Whitler}}]{Bunker2023}
{Bunker}, A.~J., {Saxena}, A., {Cameron}, A.~J., {et~al.} 2023, \aap, 677, A88,
  \dodoi{10.1051/0004-6361/202346159}

\bibitem[{{Calzetti} {et~al.}(2000){Calzetti}, {Armus}, {Bohlin}, {Kinney},
  {Koornneef}, \& {Storchi-Bergmann}}]{Calzetti2000}
{Calzetti}, D., {Armus}, L., {Bohlin}, R.~C., {et~al.} 2000, \apj, 533, 682,
  \dodoi{10.1086/308692}

\bibitem[{{Cameron} {et~al.}(2023){Cameron}, {Saxena}, {Bunker}, {D'Eugenio},
  {Carniani}, {Maiolino}, {Curtis-Lake}, {Ferruit}, {Jakobsen}, {Arribas},
  {Bonaventura}, {Charlot}, {Chevallard}, {Curti}, {Looser}, {Maseda}, {Rawle},
  {Rodr{\'\i}guez Del Pino}, {Smit}, {{\"U}bler}, {Willott}, {Witstok},
  {Egami}, {Eisenstein}, {Johnson}, {Hainline}, {Rieke}, {Robertson}, {Stark},
  {Tacchella}, {Williams}, {Willmer}, {Bhatawdekar}, {Bowler}, {Boyett},
  {Circosta}, {Helton}, {Jones}, {Kumari}, {Ji}, {Nelson}, {Parlanti},
  {Sandles}, {Scholtz}, \& {Sun}}]{Cameron2023}
{Cameron}, A.~J., {Saxena}, A., {Bunker}, A.~J., {et~al.} 2023, \aap, 677,
  A115, \dodoi{10.1051/0004-6361/202346107}

\bibitem[{{Carnall} {et~al.}(2018){Carnall}, {McLure}, {Dunlop}, \&
  {Dav{\'e}}}]{Carnall2018}
{Carnall}, A.~C., {McLure}, R.~J., {Dunlop}, J.~S., \& {Dav{\'e}}, R. 2018,
  \mnras, 480, 4379, \dodoi{10.1093/mnras/sty2169}

\bibitem[{{Carniani} {et~al.}(2020){Carniani}, {Ferrara}, {Maiolino},
  {Castellano}, {Gallerani}, {Fontana}, {Kohandel}, {Lupi}, {Pallottini},
  {Pentericci}, {Vallini}, \& {Vanzella}}]{Carniani2020}
{Carniani}, S., {Ferrara}, A., {Maiolino}, R., {et~al.} 2020, \mnras, 499,
  5136, \dodoi{10.1093/mnras/staa3178}

\bibitem[{{Carniani} {et~al.}(2024){Carniani}, {Hainline}, {D'Eugenio},
  {Eisenstein}, {Jakobsen}, {Witstok}, {Johnson}, {Chevallard}, {Maiolino},
  {Helton}, {Willott}, {Robertson}, {Alberts}, {Arribas}, {Baker},
  {Bhatawdekar}, {Boyett}, {Bunker}, {Cameron}, {Cargile}, {Charlot}, {Curti},
  {Curtis-Lake}, {Egami}, {Giardino}, {Isaak}, {Ji}, {Jones}, {Kumari},
  {Maseda}, {Parlanti}, {P{\'e}rez-Gonz{\'a}lez}, {Rawle}, {Rieke}, {Rieke},
  {Del Pino}, {Saxena}, {Scholtz}, {Smit}, {Sun}, {Tacchella}, {{\"U}bler},
  {Venturi}, {Williams}, \& {Willmer}}]{Carniani2024}
{Carniani}, S., {Hainline}, K., {D'Eugenio}, F., {et~al.} 2024, \nat, 633, 318,
  \dodoi{10.1038/s41586-024-07860-9}

\bibitem[{{Castellano} {et~al.}(2024){Castellano}, {Napolitano}, {Fontana},
  {Roberts-Borsani}, {Treu}, {Vanzella}, {Zavala}, {Arrabal Haro},
  {Calabr{\`o}}, {Llerena}, {Mascia}, {Merlin}, {Paris}, {Pentericci},
  {Santini}, {Bakx}, {Bergamini}, {Cupani}, {Dickinson}, {Filippenko},
  {Glazebrook}, {Grillo}, {Kelly}, {Malkan}, {Mason}, {Morishita},
  {Nanayakkara}, {Rosati}, {Sani}, {Wang}, \& {Yoon}}]{Castellano2024}
{Castellano}, M., {Napolitano}, L., {Fontana}, A., {et~al.} 2024, \apj, 972,
  143, \dodoi{10.3847/1538-4357/ad5f88}

\bibitem[{{Cormier} {et~al.}(2015){Cormier}, {Madden}, {Lebouteiller}, {Abel},
  {Hony}, {Galliano}, {R{\'e}my-Ruyer}, {Bigiel}, {Baes}, {Boselli},
  {Chevance}, {Cooray}, {De Looze}, {Doublier}, {Galametz}, {Hughes},
  {Karczewski}, {Lee}, {Lu}, \& {Spinoglio}}]{Cormier2015}
{Cormier}, D., {Madden}, S.~C., {Lebouteiller}, V., {et~al.} 2015, \aap, 578,
  A53, \dodoi{10.1051/0004-6361/201425207}

\bibitem[{{Cormier} {et~al.}(2019){Cormier}, {Abel}, {Hony}, {Lebouteiller},
  {Madden}, {Polles}, {Galliano}, {De Looze}, {Galametz}, \&
  {Lambert-Huyghe}}]{Cormier2019}
{Cormier}, D., {Abel}, N.~P., {Hony}, S., {et~al.} 2019, \aap, 626, A23,
  \dodoi{10.1051/0004-6361/201834457}

\bibitem[{{Curtis-Lake} {et~al.}(2023){Curtis-Lake}, {Carniani}, {Cameron},
  {Charlot}, {Jakobsen}, {Maiolino}, {Bunker}, {Witstok}, {Smit}, {Chevallard},
  {Willott}, {Ferruit}, {Arribas}, {Bonaventura}, {Curti}, {D'Eugenio},
  {Franx}, {Giardino}, {Looser}, {L{\"u}tzgendorf}, {Maseda}, {Rawle}, {Rix},
  {Rodr{\'\i}guez del Pino}, {{\"U}bler}, {Sirianni}, {Dressler}, {Egami},
  {Eisenstein}, {Endsley}, {Hainline}, {Hausen}, {Johnson}, {Rieke},
  {Robertson}, {Shivaei}, {Stark}, {Tacchella}, {Williams}, {Willmer},
  {Bhatawdekar}, {Bowler}, {Boyett}, {Chen}, {de Graaff}, {Helton}, {Hviding},
  {Jones}, {Kumari}, {Lyu}, {Nelson}, {Perna}, {Sandles}, {Saxena}, {Suess},
  {Sun}, {Topping}, {Wallace}, \& {Whitler}}]{Curtis-Lake2023}
{Curtis-Lake}, E., {Carniani}, S., {Cameron}, A., {et~al.} 2023, Nature
  Astronomy, 7, 622, \dodoi{10.1038/s41550-023-01918-w}

\bibitem[{{De Looze} {et~al.}(2011){De Looze}, {Baes}, {Bendo}, {Cortese}, \&
  {Fritz}}]{DeLooze2011}
{De Looze}, I., {Baes}, M., {Bendo}, G.~J., {Cortese}, L., \& {Fritz}, J. 2011,
  \mnras, 416, 2712, \dodoi{10.1111/j.1365-2966.2011.19223.x}

\bibitem[{{De Looze} {et~al.}(2014){De Looze}, {Cormier}, {Lebouteiller},
  {Madden}, {Baes}, {Bendo}, {Boquien}, {Boselli}, {Clements}, {Cortese},
  {Cooray}, {Galametz}, {Galliano}, {Graci{\'a}-Carpio}, {Isaak}, {Karczewski},
  {Parkin}, {Pellegrini}, {R{\'e}my-Ruyer}, {Spinoglio}, {Smith}, \&
  {Sturm}}]{DeLooze2014}
{De Looze}, I., {Cormier}, D., {Lebouteiller}, V., {et~al.} 2014, \aap, 568,
  A62, \dodoi{10.1051/0004-6361/201322489}

\bibitem[{{Desprez} {et~al.}(2024){Desprez}, {Martis}, {Asada}, {Sawicki},
  {Willott}, {Muzzin}, {Abraham}, {Brada{\v{c}}}, {Brammer},
  {Estrada-Carpenter}, {Iyer}, {Matharu}, {Mowla}, {Noirot}, {Sarrouh},
  {Strait}, {Gledhill}, \& {Rihtar{\v{s}}i{\v{c}}}}]{Desprez2024}
{Desprez}, G., {Martis}, N.~S., {Asada}, Y., {et~al.} 2024, \mnras, 530, 2935,
  \dodoi{10.1093/mnras/stae1084}

\bibitem[{{D{\'\i}az-Santos} {et~al.}(2017){D{\'\i}az-Santos}, {Armus},
  {Charmandaris}, {Lu}, {Stierwalt}, {Stacey}, {Malhotra}, {van der Werf},
  {Howell}, {Privon}, {Mazzarella}, {Goldsmith}, {Murphy}, {Barcos-Mu{\~n}oz},
  {Linden}, {Inami}, {Larson}, {Evans}, {Appleton}, {Iwasawa}, {Lord},
  {Sanders}, \& {Surace}}]{Diaz-Santos2017}
{D{\'\i}az-Santos}, T., {Armus}, L., {Charmandaris}, V., {et~al.} 2017, \apj,
  846, 32, \dodoi{10.3847/1538-4357/aa81d7}

\bibitem[{{Dopita} {et~al.}(2006){Dopita}, {Fischera}, {Sutherland}, {Kewley},
  {Leitherer}, {Tuffs}, {Popescu}, {van Breugel}, \& {Groves}}]{Dopita2006}
{Dopita}, M.~A., {Fischera}, J., {Sutherland}, R.~S., {et~al.} 2006, \apjs,
  167, 177, \dodoi{10.1086/508261}

\bibitem[{{Draine}(2011)}]{Draine2011}
{Draine}, B.~T. 2011, {Physics of the Interstellar and Intergalactic Medium}

\bibitem[{{Eldridge} \& {Stanway}(2016)}]{Eldridge2016}
{Eldridge}, J.~J., \& {Stanway}, E.~R. 2016, \mnras, 462, 3302,
  \dodoi{10.1093/mnras/stw1772}

\bibitem[{{Ferland} {et~al.}(2017){Ferland}, {Chatzikos}, {Guzm{\'a}n},
  {Lykins}, {van Hoof}, {Williams}, {Abel}, {Badnell}, {Keenan}, {Porter}, \&
  {Stancil}}]{Ferland2017}
{Ferland}, G.~J., {Chatzikos}, M., {Guzm{\'a}n}, F., {et~al.} 2017, \rmxaa, 53,
  385.
\newblock \doarXiv{1705.10877}

\bibitem[{{F{\"o}rster Schreiber} {et~al.}(2001){F{\"o}rster Schreiber},
  {Genzel}, {Lutz}, {Kunze}, \& {Sternberg}}]{Forster-Schreier2001}
{F{\"o}rster Schreiber}, N.~M., {Genzel}, R., {Lutz}, D., {Kunze}, D., \&
  {Sternberg}, A. 2001, \apj, 552, 544, \dodoi{10.1086/320546}

\bibitem[{{Fudamoto} {et~al.}(2023){Fudamoto}, {Inoue}, \&
  {Sugahara}}]{Fudamoto2023}
{Fudamoto}, Y., {Inoue}, A.~K., \& {Sugahara}, Y. 2023, \mnras, 521, 2962,
  \dodoi{10.1093/mnras/stad743}

\bibitem[{{Fudamoto} {et~al.}(2021){Fudamoto}, {Oesch}, {Schouws}, {Stefanon},
  {Smit}, {Bouwens}, {Bowler}, {Endsley}, {Gonzalez}, {Inami}, {Labbe},
  {Stark}, {Aravena}, {Barrufet}, {da Cunha}, {Dayal}, {Ferrara}, {Graziani},
  {Hodge}, {Hutter}, {Li}, {De Looze}, {Nanayakkara}, {Pallottini}, {Riechers},
  {Schneider}, {Ucci}, {van der Werf}, \& {White}}]{Fudamoto2021}
{Fudamoto}, Y., {Oesch}, P.~A., {Schouws}, S., {et~al.} 2021, \nat, 597, 489,
  \dodoi{10.1038/s41586-021-03846-z}

\bibitem[{{Fujimoto} {et~al.}(2024){Fujimoto}, {Ouchi}, {Nakajima}, {Harikane},
  {Isobe}, {Brammer}, {Oguri}, {Gim{\'e}nez-Arteaga}, {Heintz}, {Kokorev},
  {Bauer}, {Ferrara}, {Kojima}, {Lagos}, {Laura}, {Schaerer}, {Shimasaku},
  {Hatsukade}, {Kohno}, {Sun}, {Valentino}, {Watson}, {Fudamoto}, {Inoue},
  {Gonz{\'a}lez-L{\'o}pez}, {Koekemoer}, {Knudsen}, {Lee}, {Magdis}, {Richard},
  {Strait}, {Sugahara}, {Tamura}, {Toft}, {Umehata}, \& {Walth}}]{Fujimoto2024}
{Fujimoto}, S., {Ouchi}, M., {Nakajima}, K., {et~al.} 2024, \apj, 964, 146,
  \dodoi{10.3847/1538-4357/ad235c}

\bibitem[{{Grevesse} {et~al.}(2010){Grevesse}, {Asplund}, {Sauval}, \&
  {Scott}}]{Grevesse2010}
{Grevesse}, N., {Asplund}, M., {Sauval}, A.~J., \& {Scott}, P. 2010, \apss,
  328, 179, \dodoi{10.1007/s10509-010-0288-z}

\bibitem[{{Groves} {et~al.}(2004){Groves}, {Dopita}, \&
  {Sutherland}}]{Groves2004}
{Groves}, B.~A., {Dopita}, M.~A., \& {Sutherland}, R.~S. 2004, \apjs, 153, 9,
  \dodoi{10.1086/421113}

\bibitem[{{Harikane} {et~al.}(2024){Harikane}, {Nakajima}, {Ouchi}, {Umeda},
  {Isobe}, {Ono}, {Xu}, \& {Zhang}}]{Harikane2024}
{Harikane}, Y., {Nakajima}, K., {Ouchi}, M., {et~al.} 2024, \apj, 960, 56,
  \dodoi{10.3847/1538-4357/ad0b7e}

\bibitem[{{Harikane} {et~al.}(2020){Harikane}, {Ouchi}, {Inoue}, {Matsuoka},
  {Tamura}, {Bakx}, {Fujimoto}, {Moriwaki}, {Ono}, {Nagao}, {Tadaki}, {Kojima},
  {Shibuya}, {Egami}, {Ferrara}, {Gallerani}, {Hashimoto}, {Kohno}, {Matsuda},
  {Matsuo}, {Pallottini}, {Sugahara}, \& {Vallini}}]{Harikane2020}
{Harikane}, Y., {Ouchi}, M., {Inoue}, A.~K., {et~al.} 2020, \apj, 896, 93,
  \dodoi{10.3847/1538-4357/ab94bd}

\bibitem[{Harris {et~al.}(2020)Harris, Millman, van~der Walt, Gommers,
  Virtanen, Cournapeau, Wieser, Taylor, Berg, Smith, Kern, Picus, Hoyer, van
  Kerkwijk, Brett, Haldane, del R{'{\i}}o, Wiebe, Peterson,
  G{'{e}}rard-Marchant, Sheppard, Reddy, Weckesser, Abbasi, Gohlke, \&
  Oliphant}]{Harris20}
Harris, C.~R., Millman, K.~J., van~der Walt, S.~J., {et~al.} 2020, Nature, 585,
  357, \dodoi{10.1038/s41586-020-2649-2}

\bibitem[{{Harshan} {et~al.}(2024){Harshan}, {Tripodi}, {Martis},
  {Rihtar{\v{s}}i{\v{c}}}, {Brada{\v{c}}}, {Asada}, {Brammer}, {Desprez},
  {Estrada-Carpenter}, {Matharu}, {Markov}, {Muzzin}, {Mowla}, {Noirot},
  {Sarrouh}, {Sawicki}, {Strait}, \& {Willott}}]{Harshan2024}
{Harshan}, A., {Tripodi}, R., {Martis}, N.~S., {et~al.} 2024, \apjl, 977, L36,
  \dodoi{10.3847/2041-8213/ad9741}

\bibitem[{{Hashimoto} {et~al.}(2019{\natexlab{a}}){Hashimoto}, {Inoue},
  {Tamura}, {Matsuo}, {Mawatari}, \& {Yamaguchi}}]{Hashimoto2019QSO}
{Hashimoto}, T., {Inoue}, A.~K., {Tamura}, Y., {et~al.} 2019{\natexlab{a}},
  \pasj, 71, 109, \dodoi{10.1093/pasj/psz094}

\bibitem[{{Hashimoto} {et~al.}(2018){Hashimoto}, {Laporte}, {Mawatari},
  {Ellis}, {Inoue}, {Zackrisson}, {Roberts-Borsani}, {Zheng}, {Tamura},
  {Bauer}, {Fletcher}, {Harikane}, {Hatsukade}, {Hayatsu}, {Matsuda}, {Matsuo},
  {Okamoto}, {Ouchi}, {Pell{\'o}}, {Rydberg}, {Shimizu}, {Taniguchi},
  {Umehata}, \& {Yoshida}}]{Hashimoto2018}
{Hashimoto}, T., {Laporte}, N., {Mawatari}, K., {et~al.} 2018, \nat, 557, 392,
  \dodoi{10.1038/s41586-018-0117-z}

\bibitem[{{Hashimoto} {et~al.}(2019{\natexlab{b}}){Hashimoto}, {Inoue},
  {Mawatari}, {Tamura}, {Matsuo}, {Furusawa}, {Harikane}, {Shibuya}, {Knudsen},
  {Kohno}, {Ono}, {Zackrisson}, {Okamoto}, {Kashikawa}, {Oesch}, {Ouchi},
  {Ota}, {Shimizu}, {Taniguchi}, {Umehata}, \& {Watson}}]{Hashimoto2019BTD}
{Hashimoto}, T., {Inoue}, A.~K., {Mawatari}, K., {et~al.} 2019{\natexlab{b}},
  \pasj, 71, 71, \dodoi{10.1093/pasj/psz049}

\bibitem[{{Hashimoto} {et~al.}(2023){Hashimoto}, {{\'A}lvarez-M{\'a}rquez},
  {Fudamoto}, {Colina}, {Inoue}, {Nakazato}, {Ceverino}, {Yoshida},
  {Costantin}, {Sugahara}, {G{\'o}mez}, {Blanco-Prieto}, {Mawatari}, {Arribas},
  {Marques-Chaves}, {Pereira-Santaella}, {Bakx}, {Hagimoto}, {Hashigaya},
  {Matsuo}, {Tamura}, {Usui}, \& {Ren}}]{Hashimoto2023}
{Hashimoto}, T., {{\'A}lvarez-M{\'a}rquez}, J., {Fudamoto}, Y., {et~al.} 2023,
  \apjl, 955, L2, \dodoi{10.3847/2041-8213/acf57c}

\bibitem[{{Herrera-Camus} {et~al.}(2015){Herrera-Camus}, {Bolatto}, {Wolfire},
  {Smith}, {Croxall}, {Kennicutt}, {Calzetti}, {Helou}, {Walter}, {Leroy},
  {Draine}, {Brandl}, {Armus}, {Sandstrom}, {Dale}, {Aniano}, {Meidt},
  {Boquien}, {Hunt}, {Galametz}, {Tabatabaei}, {Murphy}, {Appleton}, {Roussel},
  {Engelbracht}, \& {Beirao}}]{Herrera-Camus2015}
{Herrera-Camus}, R., {Bolatto}, A.~D., {Wolfire}, M.~G., {et~al.} 2015, \apj,
  800, 1, \dodoi{10.1088/0004-637X/800/1/1}

\bibitem[{{Herrera-Camus} {et~al.}(2018){Herrera-Camus}, {Sturm},
  {Graci{\'a}-Carpio}, {Lutz}, {Contursi}, {Veilleux}, {Fischer},
  {Gonz{\'a}lez-Alfonso}, {Poglitsch}, {Tacconi}, {Genzel}, {Maiolino},
  {Sternberg}, {Davies}, \& {Verma}}]{Herrera-Camus2018}
{Herrera-Camus}, R., {Sturm}, E., {Graci{\'a}-Carpio}, J., {et~al.} 2018, \apj,
  861, 94, \dodoi{10.3847/1538-4357/aac0f6}

\bibitem[{{Herrera-Camus} {et~al.}(2025){Herrera-Camus},
  {Gonz{\'a}lez-L{\'o}pez}, {F{\"o}rster Schreiber}, {Aravena}, {de Looze},
  {Spilker}, {Tadaki}, {Barcos-Mu{\~n}oz}, {Assef}, {Birkin}, {Bolatto},
  {Bouwens}, {Bovino}, {Bowler}, {Calistro Rivera}, {da Cunha}, {Davies},
  {Davies}, {D{\'\i}az-Santos}, {Ferrara}, {Fisher}, {Genzel}, {Hodge},
  {Ikeda}, {Killi}, {Lee}, {Li}, {Li}, {Liu}, {Lutz}, {Mitsuhashi},
  {Narayanan}, {Naab}, {Palla}, {Price}, {Posses}, {Rela{\~n}o}, {Smit},
  {Solimano}, {Sternberg}, {Tacconi}, {Telikova}, {{\"U}bler}, {van der
  Giessen}, {Veilleux}, {Villanueva}, \& {Baeza-Garay}}]{Herrera-Camus2025}
{Herrera-Camus}, R., {Gonz{\'a}lez-L{\'o}pez}, J., {F{\"o}rster Schreiber}, N.,
  {et~al.} 2025, arXiv e-prints, arXiv:2505.06340.
\newblock \doarXiv{2505.06340}

\bibitem[{{Hollenbach} \& {Tielens}(1999)}]{Hollenbach1999}
{Hollenbach}, D.~J., \& {Tielens}, A.~G.~G.~M. 1999, Reviews of Modern Physics,
  71, 173, \dodoi{10.1103/RevModPhys.71.173}

\bibitem[{Hoyer \& Hamman(2016)}]{Hoyer17}
Hoyer, S., \& Hamman, J. 2016, Journal of Open Research Software, 5,
  \dodoi{10.5334/jors.148}

\bibitem[{{Hughes} {et~al.}(2020){Hughes}, {Schloerb}, {Aretxaga},
  {Castillo-Dom{\'\i}nguez}, {Ch{\'a}vez Dagostino}, {Col{\'\i}n}, {Erickson},
  {Ferrusca Rodriguez}, {Gale}, {G{\'o}mez-Ruiz}, {Hern{\'a}ndez Rebollar},
  {Heyer}, {Lowenthal}, {Monta{\~n}a}, {Moreno Nolasco}, {Narayanan}, {Pope},
  {Rodr{\'\i}guez-Montoya}, {S{\'a}nchez-Arg{\"u}elles}, {Smith}, {Souccar},
  {de la Rosa Becerra}, {Wilson}, \& {Yun}}]{Hughes2020}
{Hughes}, D.~H., {Schloerb}, F.~P., {Aretxaga}, I., {et~al.} 2020, in Society
  of Photo-Optical Instrumentation Engineers (SPIE) Conference Series, Vol.
  11445, Ground-based and Airborne Telescopes VIII, ed. H.~K. {Marshall},
  J.~{Spyromilio}, \& T.~{Usuda}, 1144522, \dodoi{10.1117/12.2561893}

\bibitem[{Hunter(2007)}]{Hunter07}
Hunter, J.~D. 2007, Computing in Science \& Engineering, 9, 90,
  \dodoi{10.1109/MCSE.2007.55}

\bibitem[{{Iani} {et~al.}(2023){Iani}, {Annunziatella}, {Bartosch Caminha},
  {Caputi}, {Costantin}, {Kerutt}, {Kokorev}, {Navarro}, {Perez-Gonzalez},
  {Rinaldi}, {Yang}, \& {van Mierlo}}]{Iani2023}
{Iani}, E., {Annunziatella}, M., {Bartosch Caminha}, G., {et~al.} 2023,
  {Unveiling the properties of high-redshift low/intermediate-mass galaxies in
  Lensing fields with NIRCam Wide Field Slitless Spectroscopy}, JWST Proposal.
  Cycle 2, ID. \#3538

\bibitem[{{Inami} {et~al.}(2022){Inami}, {Algera}, {Schouws}, {Sommovigo},
  {Bouwens}, {Smit}, {Stefanon}, {Bowler}, {Endsley}, {Ferrara}, {Oesch},
  {Stark}, {Aravena}, {Barrufet}, {da Cunha}, {Dayal}, {De Looze}, {Fudamoto},
  {Gonzalez}, {Graziani}, {Hodge}, {Hygate}, {Nanayakkara}, {Pallottini},
  {Riechers}, {Schneider}, {Topping}, \& {van der Werf}}]{Inami2022}
{Inami}, H., {Algera}, H. S.~B., {Schouws}, S., {et~al.} 2022, \mnras, 515,
  3126, \dodoi{10.1093/mnras/stac1779}

\bibitem[{{Indriolo} {et~al.}(2007){Indriolo}, {Geballe}, {Oka}, \&
  {McCall}}]{Indriolo2007}
{Indriolo}, N., {Geballe}, T.~R., {Oka}, T., \& {McCall}, B.~J. 2007, \apj,
  671, 1736, \dodoi{10.1086/523036}

\bibitem[{{Infante} {et~al.}(2015){Infante}, {Zheng}, {Laporte}, {Troncoso
  Iribarren}, {Molino}, {Diego}, {Bauer}, {Zitrin}, {Moustakas}, {Huang},
  {Shu}, {Bina}, {Brammer}, {Broadhurst}, {Ford}, {Garc{\'\i}a}, \&
  {Kim}}]{Infante2015}
{Infante}, L., {Zheng}, W., {Laporte}, N., {et~al.} 2015, \apj, 815, 18,
  \dodoi{10.1088/0004-637X/815/1/18}

\bibitem[{{Inoue} {et~al.}(2014){Inoue}, {Shimizu}, {Tamura}, {Matsuo},
  {Okamoto}, \& {Yoshida}}]{Inoue2014}
{Inoue}, A.~K., {Shimizu}, I., {Tamura}, Y., {et~al.} 2014, \apjl, 780, L18,
  \dodoi{10.1088/2041-8205/780/2/L18}

\bibitem[{{Inoue} {et~al.}(2016){Inoue}, {Tamura}, {Matsuo}, {Mawatari},
  {Shimizu}, {Shibuya}, {Ota}, {Yoshida}, {Zackrisson}, {Kashikawa}, {Kohno},
  {Umehata}, {Hatsukade}, {Iye}, {Matsuda}, {Okamoto}, \&
  {Yamaguchi}}]{Inoue2016}
{Inoue}, A.~K., {Tamura}, Y., {Matsuo}, H., {et~al.} 2016, Science, 352, 1559,
  \dodoi{10.1126/science.aaf0714}

\bibitem[{{Iyer} {et~al.}(2019){Iyer}, {Gawiser}, {Faber}, {Ferguson},
  {Kartaltepe}, {Koekemoer}, {Pacifici}, \& {Somerville}}]{Iyer2019}
{Iyer}, K.~G., {Gawiser}, E., {Faber}, S.~M., {et~al.} 2019, \apj, 879, 116,
  \dodoi{10.3847/1538-4357/ab2052}

\bibitem[{{Izotov} {et~al.}(2006){Izotov}, {Stasi{\'n}ska}, {Meynet}, {Guseva},
  \& {Thuan}}]{Izotov2006}
{Izotov}, Y.~I., {Stasi{\'n}ska}, G., {Meynet}, G., {Guseva}, N.~G., \&
  {Thuan}, T.~X. 2006, \aap, 448, 955, \dodoi{10.1051/0004-6361:20053763}

\bibitem[{{Jones} {et~al.}(2024{\natexlab{a}}){Jones}, {Witstok}, {Concas}, \&
  {Laporte}}]{Jones2024_Y1}
{Jones}, G.~C., {Witstok}, J., {Concas}, A., \& {Laporte}, N.
  2024{\natexlab{a}}, \mnras, 529, L1, \dodoi{10.1093/mnrasl/slad189}

\bibitem[{{Jones} {et~al.}(2024{\natexlab{b}}){Jones}, {{\"U}bler}, {Perna},
  {Arribas}, {Bunker}, {Carniani}, {Charlot}, {Maiolino}, {Del Pino},
  {Willott}, {Bowler}, {B{\"o}ker}, {Cameron}, {Chevallard}, {Cresci}, {Curti},
  {D'Eugenio}, {Kumari}, {Saxena}, {Scholtz}, {Venturi}, \&
  {Witstok}}]{Jones2024_HFLS3}
{Jones}, G.~C., {{\"U}bler}, H., {Perna}, M., {et~al.} 2024{\natexlab{b}},
  \aap, 682, A122, \dodoi{10.1051/0004-6361/202347838}

\bibitem[{{Jones} {et~al.}(2023){Jones}, {Sanders}, {Chen}, {Wang},
  {Morishita}, {Roberts-Borsani}, {Treu}, {Dressler}, {Merlin}, {Paris},
  {Santini}, {Bergamini}, {Henry}, {Huntzinger}, {Nanayakkara}, {Boyett},
  {Bradac}, {Brammer}, {Calabr{\'o}}, {Glazebrook}, {Grasha}, {Mascia},
  {Pentericci}, {Trenti}, \& {Vulcani}}]{Jones2023}
{Jones}, T., {Sanders}, R., {Chen}, Y., {et~al.} 2023, \apjl, 951, L17,
  \dodoi{10.3847/2041-8213/acd938}

\bibitem[{{Katz} {et~al.}(2022){Katz}, {Rosdahl}, {Kimm}, {Garel}, {Blaizot},
  {Haehnelt}, {Michel-Dansac}, {Martin-Alvarez}, {Devriendt}, {Slyz},
  {Teyssier}, {Ocvirk}, {Laporte}, \& {Ellis}}]{Katz2022}
{Katz}, H., {Rosdahl}, J., {Kimm}, T., {et~al.} 2022, \mnras, 510, 5603,
  \dodoi{10.1093/mnras/stac028}

\bibitem[{{Kawamata} {et~al.}(2016){Kawamata}, {Oguri}, {Ishigaki},
  {Shimasaku}, \& {Ouchi}}]{Kawamata2016}
{Kawamata}, R., {Oguri}, M., {Ishigaki}, M., {Shimasaku}, K., \& {Ouchi}, M.
  2016, \apj, 819, 114, \dodoi{10.3847/0004-637X/819/2/114}

\bibitem[{{Kewley} \& {Dopita}(2002)}]{Kewley-Dopita2002}
{Kewley}, L.~J., \& {Dopita}, M.~A. 2002, \apjs, 142, 35,
  \dodoi{10.1086/341326}

\bibitem[{{Kewley} {et~al.}(2019){Kewley}, {Nicholls}, \&
  {Sutherland}}]{Kewley2019}
{Kewley}, L.~J., {Nicholls}, D.~C., \& {Sutherland}, R.~S. 2019, \araa, 57,
  511, \dodoi{10.1146/annurev-astro-081817-051832}

\bibitem[{{Killi} {et~al.}(2023){Killi}, {Watson}, {Fujimoto}, {Akins},
  {Knudsen}, {Richard}, {Harikane}, {Rigopoulou}, {Rizzo}, {Ginolfi},
  {Popping}, \& {Kokorev}}]{Killi2023}
{Killi}, M., {Watson}, D., {Fujimoto}, S., {et~al.} 2023, \mnras, 521, 2526,
  \dodoi{10.1093/mnras/stad687}

\bibitem[{{Kim} \& {Koo}(2001)}]{Kim2001}
{Kim}, K.-T., \& {Koo}, B.-C. 2001, \apj, 549, 979, \dodoi{10.1086/319447}

\bibitem[{{Kumari} {et~al.}(2024){Kumari}, {Smit}, {Leitherer}, {Witstok},
  {Irwin}, {Sirianni}, \& {Aloisi}}]{Kumari2024MNRAS}
{Kumari}, N., {Smit}, R., {Leitherer}, C., {et~al.} 2024, \mnras, 529, 781,
  \dodoi{10.1093/mnras/stae252}

\bibitem[{{Kurtz}(2005)}]{Kurtz2005}
{Kurtz}, S. 2005, in Massive Star Birth: A Crossroads of Astrophysics, ed.
  R.~{Cesaroni}, M.~{Felli}, E.~{Churchwell}, \& M.~{Walmsley}, Vol. 227,
  111--119, \dodoi{10.1017/S1743921305004424}

\bibitem[{{Laporte} {et~al.}(2015){Laporte}, {Streblyanska}, {Kim},
  {Pell{\'o}}, {Bauer}, {Bina}, {Brammer}, {De Leo}, {Infante}, \&
  {P{\'e}rez-Fournon}}]{Laporte2015}
{Laporte}, N., {Streblyanska}, A., {Kim}, S., {et~al.} 2015, \aap, 575, A92,
  \dodoi{10.1051/0004-6361/201425040}

\bibitem[{{Laporte} {et~al.}(2019){Laporte}, {Katz}, {Ellis}, {Lagache},
  {Bauer}, {Boone}, {Inoue}, {Hashimoto}, {Matsuo}, {Mawatari}, \&
  {Tamura}}]{Laporte2019}
{Laporte}, N., {Katz}, H., {Ellis}, R.~S., {et~al.} 2019, \mnras, 487, L81,
  \dodoi{10.1093/mnrasl/slz094}

\bibitem[{{Le F{\`e}vre} {et~al.}(2020){Le F{\`e}vre}, {B{\'e}thermin},
  {Faisst}, {Jones}, {Capak}, {Cassata}, {Silverman}, {Schaerer}, {Yan},
  {Amorin}, {Bardelli}, {Boquien}, {Cimatti}, {Dessauges-Zavadsky},
  {Giavalisco}, {Hathi}, {Fudamoto}, {Fujimoto}, {Ginolfi}, {Gruppioni},
  {Hemmati}, {Ibar}, {Koekemoer}, {Khusanova}, {Lagache}, {Lemaux}, {Loiacono},
  {Maiolino}, {Mancini}, {Narayanan}, {Morselli}, {M{\'e}ndez-Hern{\`a}ndez},
  {Oesch}, {Pozzi}, {Romano}, {Riechers}, {Scoville}, {Talia}, {Tasca},
  {Thomas}, {Toft}, {Vallini}, {Vergani}, {Walter}, {Zamorani}, \&
  {Zucca}}]{LeFerve2020}
{Le F{\`e}vre}, O., {B{\'e}thermin}, M., {Faisst}, A., {et~al.} 2020, \aap,
  643, A1, \dodoi{10.1051/0004-6361/201936965}

\bibitem[{{Liang} {et~al.}(2024){Liang}, {Feldmann}, {Murray}, {Narayanan},
  {Hayward}, {Angl{\'e}s-Alc{\'a}zar}, {Bassini}, {Richings},
  {Faucher-Gigu{\`e}re}, {Chung}, {Chan}, {Tolgay}, {{\c{C}}atmabacak},
  {Kere{\v{s}}}, \& {Hopkins}}]{Liang2024}
{Liang}, L., {Feldmann}, R., {Murray}, N., {et~al.} 2024, \mnras, 528, 499,
  \dodoi{10.1093/mnras/stad3792}

\bibitem[{{Luridiana} {et~al.}(2015){Luridiana}, {Morisset}, \&
  {Shaw}}]{Luridiana2015}
{Luridiana}, V., {Morisset}, C., \& {Shaw}, R.~A. 2015, \aap, 573, A42,
  \dodoi{10.1051/0004-6361/201323152}

\bibitem[{{Ma} {et~al.}(2024){Ma}, {Sun}, {Cheng}, {Yan}, {Ling}, {Sun}, {Foo},
  {Egami}, {Diego}, {Cohen}, {Jansen}, {Summers}, {Windhorst}, {D'Silva},
  {Koekemoer}, {Coe}, {Conselice}, {Driver}, {Frye}, {Grogin}, {Marshall},
  {Nonino}, {Ortiz}, {Pirzkal}, {Robotham}, {Ryan}, {Willmer}, {Adams},
  {Hathi}, {Dole}, {Willner}, {Espada}, {Furtak}, {Hsiao}, {Li}, {Chen},
  {Jolly}, \& {Chen}}]{Ma2024}
{Ma}, Z., {Sun}, B., {Cheng}, C., {et~al.} 2024, \apj, 975, 87,
  \dodoi{10.3847/1538-4357/ad7b32}

\bibitem[{{Madau} \& {Dickinson}(2014)}]{Madau2014}
{Madau}, P., \& {Dickinson}, M. 2014, \araa, 52, 415,
  \dodoi{10.1146/annurev-astro-081811-125615}

\bibitem[{{Madden} {et~al.}(2013){Madden}, {R{\'e}my-Ruyer}, {Galametz},
  {Cormier}, {Lebouteiller}, {Galliano}, {Hony}, {Bendo}, {Smith}, {Pohlen},
  {Roussel}, {Sauvage}, {Wu}, {Sturm}, {Poglitsch}, {Contursi}, {Doublier},
  {Baes}, {Barlow}, {Boselli}, {Boquien}, {Carlson}, {Ciesla}, {Cooray},
  {Cortese}, {de Looze}, {Irwin}, {Isaak}, {Kamenetzky}, {Karczewski}, {Lu},
  {MacHattie}, {O'Halloran}, {Parkin}, {Rangwala}, {Schirm}, {Schulz},
  {Spinoglio}, {Vaccari}, {Wilson}, \& {Wozniak}}]{Madden2013}
{Madden}, S.~C., {R{\'e}my-Ruyer}, A., {Galametz}, M., {et~al.} 2013, \pasp,
  125, 600, \dodoi{10.1086/671138}

\bibitem[{{Markov} {et~al.}(2023){Markov}, {Gallerani}, {Pallottini},
  {Sommovigo}, {Carniani}, {Ferrara}, {Parlanti}, \& {Di Mascia}}]{Markov2023}
{Markov}, V., {Gallerani}, S., {Pallottini}, A., {et~al.} 2023, \aap, 679, A12,
  \dodoi{10.1051/0004-6361/202346723}

\bibitem[{{Marrone} {et~al.}(2018){Marrone}, {Spilker}, {Hayward}, {Vieira},
  {Aravena}, {Ashby}, {Bayliss}, {B{\'e}thermin}, {Brodwin}, {Bothwell},
  {Carlstrom}, {Chapman}, {Chen}, {Crawford}, {Cunningham}, {De Breuck},
  {Fassnacht}, {Gonzalez}, {Greve}, {Hezaveh}, {Lacaille}, {Litke}, {Lower},
  {Ma}, {Malkan}, {Miller}, {Morningstar}, {Murphy}, {Narayanan}, {Phadke},
  {Rotermund}, {Sreevani}, {Stalder}, {Stark}, {Strandet}, {Tang}, \&
  {Wei{\ss}}}]{Marrone2018}
{Marrone}, D.~P., {Spilker}, J.~S., {Hayward}, C.~C., {et~al.} 2018, \nat, 553,
  51, \dodoi{10.1038/nature24629}

\bibitem[{{Matsuo} {et~al.}(2009){Matsuo}, {Arai}, {Nitta}, \&
  {Kosaka}}]{Matuo2009}
{Matsuo}, H., {Arai}, T., {Nitta}, T., \& {Kosaka}, A. 2009, in Astronomical
  Society of the Pacific Conference Series, Vol. 418, AKARI, a Light to
  Illuminate the Misty Universe, ed. T.~{Onaka}, G.~J. {White}, T.~{Nakagawa},
  \& I.~{Yamamura}, 451

\bibitem[{{Mizutani} {et~al.}(2002){Mizutani}, {Onaka}, \&
  {Shibai}}]{Mizutani2002}
{Mizutani}, M., {Onaka}, T., \& {Shibai}, H. 2002, \aap, 382, 610,
  \dodoi{10.1051/0004-6361:20011611}

\bibitem[{{Morishita} {et~al.}(2023){Morishita}, {Roberts-Borsani}, {Treu},
  {Brammer}, {Mason}, {Trenti}, {Vulcani}, {Wang}, {Acebron}, {Bah{\'e}},
  {Bergamini}, {Boyett}, {Bradac}, {Calabr{\`o}}, {Castellano}, {Chen}, {De
  Lucia}, {Filippenko}, {Fontana}, {Glazebrook}, {Grillo}, {Henry}, {Jones},
  {Kelly}, {Koekemoer}, {Leethochawalit}, {Lu}, {Marchesini}, {Mascia},
  {Mercurio}, {Merlin}, {Metha}, {Nanayakkara}, {Nonino}, {Paris},
  {Pentericci}, {Rosati}, {Santini}, {Strait}, {Vanzella}, {Windhorst}, \&
  {Xie}}]{Morishita2023}
{Morishita}, T., {Roberts-Borsani}, G., {Treu}, T., {et~al.} 2023, \apjl, 947,
  L24, \dodoi{10.3847/2041-8213/acb99e}

\bibitem[{{Nagao} {et~al.}(2011){Nagao}, {Maiolino}, {Marconi}, \&
  {Matsuhara}}]{Nagao2011}
{Nagao}, T., {Maiolino}, R., {Marconi}, A., \& {Matsuhara}, H. 2011, \aap, 526,
  A149, \dodoi{10.1051/0004-6361/201015471}

\bibitem[{{Naidu} {et~al.}(2020){Naidu}, {Tacchella}, {Mason}, {Bose}, {Oesch},
  \& {Conroy}}]{Naidu2020}
{Naidu}, R.~P., {Tacchella}, S., {Mason}, C.~A., {et~al.} 2020, \apj, 892, 109,
  \dodoi{10.3847/1538-4357/ab7cc9}

\bibitem[{{Nakajima} \& {Ouchi}(2014)}]{Nakajima-Ouchi2014}
{Nakajima}, K., \& {Ouchi}, M. 2014, \mnras, 442, 900,
  \dodoi{10.1093/mnras/stu902}

\bibitem[{{Nakane} {et~al.}(2024){Nakane}, {Ouchi}, {Nakajima}, {Harikane},
  {Ono}, {Umeda}, {Isobe}, {Zhang}, \& {Xu}}]{Nakane2024}
{Nakane}, M., {Ouchi}, M., {Nakajima}, K., {et~al.} 2024, \apj, 967, 28,
  \dodoi{10.3847/1538-4357/ad38c2}

\bibitem[{{Nicholls} {et~al.}(2020){Nicholls}, {Kewley}, \&
  {Sutherland}}]{Nicholles2020}
{Nicholls}, D.~C., {Kewley}, L.~J., \& {Sutherland}, R.~S. 2020, \pasp, 132,
  033001, \dodoi{10.1088/1538-3873/ab6818}

\bibitem[{{Nyhagen} {et~al.}(2024){Nyhagen}, {Schimek}, {Cicone}, {Decataldo},
  \& {Shen}}]{Nyhagen2024}
{Nyhagen}, C.~T., {Schimek}, A., {Cicone}, C., {Decataldo}, D., \& {Shen}, S.
  2024, arXiv e-prints, arXiv:2410.18471, \dodoi{10.48550/arXiv.2410.18471}

\bibitem[{{Osterbrock} \& {Ferland}(2006)}]{Osterbrock2006}
{Osterbrock}, D.~E., \& {Ferland}, G.~J. 2006, {Astrophysics of gaseous nebulae
  and active galactic nuclei}

\bibitem[{{Planck Collaboration VI}(2020)}]{Planck2018parameters}
{Planck Collaboration VI}. 2020, \aap, 641, A6,
  \dodoi{10.1051/0004-6361/201833910}

\bibitem[{{Ramambason} {et~al.}(2022){Ramambason}, {Lebouteiller}, {Bik},
  {Richardson}, {Galliano}, {Schaerer}, {Morisset}, {Polles}, {Madden},
  {Chevance}, \& {De Looze}}]{Ramambason2022}
{Ramambason}, L., {Lebouteiller}, V., {Bik}, A., {et~al.} 2022, \aap, 667, A35,
  \dodoi{10.1051/0004-6361/202243866}

\bibitem[{{Rigby} \& {Rieke}(2004)}]{Rigby2004}
{Rigby}, J.~R., \& {Rieke}, G.~H. 2004, \apj, 606, 237, \dodoi{10.1086/382776}

\bibitem[{{Schaerer} {et~al.}(2022){Schaerer}, {Marques-Chaves}, {Barrufet},
  {Oesch}, {Izotov}, {Naidu}, {Guseva}, \& {Brammer}}]{Schaerer2022}
{Schaerer}, D., {Marques-Chaves}, R., {Barrufet}, L., {et~al.} 2022, \aap, 665,
  L4, \dodoi{10.1051/0004-6361/202244556}

\bibitem[{{Schaerer} {et~al.}(2020){Schaerer}, {Ginolfi}, {B{\'e}thermin},
  {Fudamoto}, {Oesch}, {Le F{\`e}vre}, {Faisst}, {Capak}, {Cassata},
  {Silverman}, {Yan}, {Jones}, {Amorin}, {Bardelli}, {Boquien}, {Cimatti},
  {Dessauges-Zavadsky}, {Giavalisco}, {Hathi}, {Fujimoto}, {Ibar}, {Koekemoer},
  {Lagache}, {Lemaux}, {Loiacono}, {Maiolino}, {Narayanan}, {Morselli},
  {M{\'e}ndez-Hern{\`a}ndez}, {Pozzi}, {Riechers}, {Talia}, {Toft}, {Vallini},
  {Vergani}, {Zamorani}, \& {Zucca}}]{Schaerer2020}
{Schaerer}, D., {Ginolfi}, M., {B{\'e}thermin}, M., {et~al.} 2020, \aap, 643,
  A3, \dodoi{10.1051/0004-6361/202037617}

\bibitem[{{Schimek} {et~al.}(2024){Schimek}, {Cicone}, {Shen}, {Decataldo},
  {Klaassen}, \& {Mayer}}]{Schimek2024}
{Schimek}, A., {Cicone}, C., {Shen}, S., {et~al.} 2024, \aap, 687, L10,
  \dodoi{10.1051/0004-6361/202449903}

\bibitem[{{Smith} {et~al.}(2006){Smith}, {Westmoquette}, {Gallagher},
  {O'Connell}, {Rosario}, \& {de Grijs}}]{Smith2006}
{Smith}, L.~J., {Westmoquette}, M.~S., {Gallagher}, J.~S., {et~al.} 2006,
  \mnras, 370, 513, \dodoi{10.1111/j.1365-2966.2006.10507.x}

\bibitem[{{Sommovigo} {et~al.}(2022){Sommovigo}, {Ferrara}, {Pallottini},
  {Dayal}, {Bouwens}, {Smit}, {da Cunha}, {De Looze}, {Bowler}, {Hodge},
  {Inami}, {Oesch}, {Endsley}, {Gonzalez}, {Schouws}, {Stark}, {Stefanon},
  {Aravena}, {Graziani}, {Riechers}, {Schneider}, {van der Werf}, {Algera},
  {Barrufet}, {Fudamoto}, {Hygate}, {Labb{\'e}}, {Li}, {Nanayakkara}, \&
  {Topping}}]{Sommovigo2022}
{Sommovigo}, L., {Ferrara}, A., {Pallottini}, A., {et~al.} 2022, \mnras, 513,
  3122, \dodoi{10.1093/mnras/stac302}

\bibitem[{{Stanway} {et~al.}(2016){Stanway}, {Eldridge}, \&
  {Becker}}]{Stanway2016}
{Stanway}, E.~R., {Eldridge}, J.~J., \& {Becker}, G.~D. 2016, \mnras, 456, 485,
  \dodoi{10.1093/mnras/stv2661}

\bibitem[{{Sugahara} {et~al.}(2022){Sugahara}, {Inoue}, {Fudamoto},
  {Hashimoto}, {Harikane}, \& {Yamanaka}}]{Sugahara2022}
{Sugahara}, Y., {Inoue}, A.~K., {Fudamoto}, Y., {et~al.} 2022, \apj, 935, 119,
  \dodoi{10.3847/1538-4357/ac7fed}

\bibitem[{{Sugahara} {et~al.}(2021){Sugahara}, {Inoue}, {Hashimoto},
  {Yamanaka}, {Fujimoto}, {Tamura}, {Matsuo}, {Binggeli}, \&
  {Zackrisson}}]{Sugahara2021}
{Sugahara}, Y., {Inoue}, A.~K., {Hashimoto}, T., {et~al.} 2021, \apj, 923, 5,
  \dodoi{10.3847/1538-4357/ac2a36}

\bibitem[{{Sugahara} {et~al.}(2025){Sugahara}, {{\'A}lvarez-M{\'a}rquez},
  {Hashimoto}, {Colina}, {Inoue}, {Costantin}, {Fudamoto}, {Mawatari}, {Ren},
  {Arribas}, {Bakx}, {Blanco-Prieto}, {Ceverino}, {Crespo G{\'o}mez},
  {Hagimoto}, {Hashigaya}, {Marques-Chaves}, {Matsuo}, {Nakazato},
  {Pereira-Santaella}, {Tamura}, {Usui}, \& {Yoshida}}]{Sugahara2024}
{Sugahara}, Y., {{\'A}lvarez-M{\'a}rquez}, J., {Hashimoto}, T., {et~al.} 2025,
  \apj, 981, 135, \dodoi{10.3847/1538-4357/adb02a}

\bibitem[{{Tadaki} {et~al.}(2022){Tadaki}, {Tsujita}, {Tamura}, {Kohno},
  {Hatsukade}, {Iono}, {Lee}, {Matsuda}, {Michiyama}, {Nagao}, {Nakanishi},
  {Nishimura}, {Saito}, {Umehata}, \& {Zavala}}]{Tadaki2022}
{Tadaki}, K.-i., {Tsujita}, A., {Tamura}, Y., {et~al.} 2022, \pasj, 74, L9,
  \dodoi{10.1093/pasj/psac018}

\bibitem[{{Takami} {et~al.}(1987){Takami}, {Maihara}, {Mizutani}, {Okuda},
  {Shibai}, {Nakagawa}, {Thomas}, {Sood}, {Hiromoto}, \&
  {Kobayashi}}]{Takami1987}
{Takami}, H., {Maihara}, T., {Mizutani}, K., {et~al.} 1987, \pasp, 99, 832,
  \dodoi{10.1086/132043}

\bibitem[{{Tamura} {et~al.}(2019){Tamura}, {Mawatari}, {Hashimoto}, {Inoue},
  {Zackrisson}, {Christensen}, {Binggeli}, {Matsuda}, {Matsuo}, {Takeuchi},
  {Asano}, {Sunaga}, {Shimizu}, {Okamoto}, {Yoshida}, {Lee}, {Shibuya},
  {Taniguchi}, {Umehata}, {Hatsukade}, {Kohno}, \& {Ota}}]{Tamura2019}
{Tamura}, Y., {Mawatari}, K., {Hashimoto}, T., {et~al.} 2019, \apj, 874, 27,
  \dodoi{10.3847/1538-4357/ab0374}

\bibitem[{{Tamura} {et~al.}(2023){Tamura}, {C. Bakx}, {Inoue}, {Hashimoto},
  {Tokuoka}, {Imamura}, {Hatsukade}, {Lee}, {Moriwaki}, {Okamoto}, {Ota},
  {Umehata}, {Yoshida}, {Zackrisson}, {Hagimoto}, {Matsuo}, {Shimizu},
  {Sugahara}, \& {Takeuchi}}]{Tamura2023}
{Tamura}, Y., {C. Bakx}, T. J.~L., {Inoue}, A.~K., {et~al.} 2023, \apj, 952, 9,
  \dodoi{10.3847/1538-4357/acd637}

\bibitem[{{Tamura} {et~al.}(2024){Tamura}, {Sakai}, {Kawabe}, {Kojima},
  {Taniguchi}, {Takekoshi}, {Kang}, {Shan}, {Hagimoto}, {Okauchi}, {Tetsuka},
  {Inoue}, {Kohno}, {Tanaka}, {Bakx}, {Fudamoto}, {Fujita}, {Harikane},
  {Hashimoto}, {Hatsukade}, {Hughes}, {Iino}, {Kimura}, {Maezawa}, {Matsuda},
  {Mawatari}, {Nakajima}, {Nakatsubo}, {Oshima}, {Sagawa}, {Schloerb},
  {Takahashi}, {Taniguchi}, {Tsujita}, {Umehata}, {Yonetsu}, \&
  {Yun}}]{Tamura2024FINER}
{Tamura}, Y., {Sakai}, T., {Kawabe}, R., {et~al.} 2024, in Society of
  Photo-Optical Instrumentation Engineers (SPIE) Conference Series, Vol. 13102,
  Society of Photo-Optical Instrumentation Engineers (SPIE) Conference Series,
  ed. J.~{Zmuidzinas} \& J.-R. {Gao}, 131020G, \dodoi{10.1117/12.3017788}

\bibitem[{{Umeda} {et~al.}(2024){Umeda}, {Ouchi}, {Nakajima}, {Harikane},
  {Ono}, {Xu}, {Isobe}, \& {Zhang}}]{Umeda2023}
{Umeda}, H., {Ouchi}, M., {Nakajima}, K., {et~al.} 2024, \apj, 971, 124,
  \dodoi{10.3847/1538-4357/ad554e}

\bibitem[{{Ura} {et~al.}(2023){Ura}, {Hashimoto}, {Inoue}, {Fadda}, {Hayes},
  {Puschnig}, {Zackrisson}, {Tamura}, {Matsuo}, {Mawatari}, {Fudamoto},
  {Hagimoto}, {Kuno}, {Sugahara}, {Yamanaka}, {Bakx}, {Nakazato}, {Usui},
  {Yajima}, \& {Yoshida}}]{Ura2023}
{Ura}, R., {Hashimoto}, T., {Inoue}, A.~K., {et~al.} 2023, \apj, 948, 3,
  \dodoi{10.3847/1538-4357/acc530}

\bibitem[{{Urquhart} {et~al.}(2013){Urquhart}, {Thompson}, {Moore}, {Purcell},
  {Hoare}, {Schuller}, {Wyrowski}, {Csengeri}, {Menten}, {Lumsden}, {Kurtz},
  {Walmsley}, {Bronfman}, {Morgan}, {Eden}, \& {Russeil}}]{Urquhart2013}
{Urquhart}, J.~S., {Thompson}, M.~A., {Moore}, T.~J.~T., {et~al.} 2013, \mnras,
  435, 400, \dodoi{10.1093/mnras/stt1310}

\bibitem[{{Vallini} {et~al.}(2021){Vallini}, {Ferrara}, {Pallottini},
  {Carniani}, \& {Gallerani}}]{Vallini2021}
{Vallini}, L., {Ferrara}, A., {Pallottini}, A., {Carniani}, S., \& {Gallerani},
  S. 2021, \mnras, 505, 5543, \dodoi{10.1093/mnras/stab1674}

\bibitem[{{Virtanen} {et~al.}(2020){Virtanen}, {Gommers}, {Oliphant},
  {Haberland}, {Reddy}, {Cournapeau}, {Burovski}, {Peterson}, {Weckesser},
  {Bright}, {van der Walt}, {Brett}, {Wilson}, {Millman}, {Mayorov}, {Nelson},
  {Jones}, {Kern}, {Larson}, {Carey}, {Polat}, {Feng}, {Moore}, {VanderPlas},
  {Laxalde}, {Perktold}, {Cimrman}, {Henriksen}, {Quintero}, {Harris},
  {Archibald}, {Ribeiro}, {Pedregosa}, {van Mulbregt}, \& {SciPy 1. 0
  Contributors}}]{Virtanen2020}
{Virtanen}, P., {Gommers}, R., {Oliphant}, T.~E., {et~al.} 2020, Nature
  Methods, 17, 261, \dodoi{10.1038/s41592-019-0686-2}

\bibitem[{{Walter} {et~al.}(2018){Walter}, {Riechers}, {Novak}, {Decarli},
  {Ferkinhoff}, {Venemans}, {Ba{\~n}ados}, {Bertoldi}, {Carilli}, {Fan},
  {Farina}, {Mazzucchelli}, {Neeleman}, {Rix}, {Strauss}, {Uzgil}, \&
  {Wang}}]{Walter2018}
{Walter}, F., {Riechers}, D., {Novak}, M., {et~al.} 2018, \apjl, 869, L22,
  \dodoi{10.3847/2041-8213/aaf4fa}

\bibitem[{{W}es {M}c{K}inney(2010)}]{mckinney-proc-scipy-2010}
{W}es {M}c{K}inney. 2010, in {P}roceedings of the 9th {P}ython in {S}cience
  {C}onference, ed. {S}t\'efan van~der {W}alt \& {J}arrod {M}illman, 56 -- 61,
  \dodoi{10.25080/Majora-92bf1922-00a}

\bibitem[{{Willott} {et~al.}(2022){Willott}, {Doyon}, {Albert}, {Brammer},
  {Dixon}, {Muzic}, {Ravindranath}, {Scholz}, {Abraham}, {Artigau},
  {Brada{\v{c}}}, {Goudfrooij}, {Hutchings}, {Iyer}, {Jayawardhana}, {LaMassa},
  {Martis}, {Meyer}, {Morishita}, {Mowla}, {Muzzin}, {Noirot}, {Pacifici},
  {Rowlands}, {Sarrouh}, {Sawicki}, {Taylor}, {Volk}, \& {Zabl}}]{Willott2022}
{Willott}, C.~J., {Doyon}, R., {Albert}, L., {et~al.} 2022, \pasp, 134, 025002,
  \dodoi{10.1088/1538-3873/ac5158}

\bibitem[{{Windhorst} {et~al.}(2023){Windhorst}, {Cohen}, {Jansen}, {Summers},
  {Tompkins}, {Conselice}, {Driver}, {Yan}, {Coe}, {Frye}, {Grogin},
  {Koekemoer}, {Marshall}, {O'Brien}, {Pirzkal}, {Robotham}, {Ryan}, {Willmer},
  {Carleton}, {Diego}, {Keel}, {Porto}, {Redshaw}, {Scheller}, {Wilkins},
  {Willner}, {Zitrin}, {Adams}, {Austin}, {Arendt}, {Beacom}, {Bhatawdekar},
  {Bradley}, {Broadhurst}, {Cheng}, {Civano}, {Dai}, {Dole}, {D'Silva},
  {Duncan}, {Fazio}, {Ferrami}, {Ferreira}, {Finkelstein}, {Furtak}, {Gim},
  {Griffiths}, {Hammel}, {Harrington}, {Hathi}, {Holwerda}, {Honor}, {Huang},
  {Hyun}, {Im}, {Joshi}, {Kamieneski}, {Kelly}, {Larson}, {Li}, {Lim}, {Ma},
  {Maksym}, {Manzoni}, {Meena}, {Milam}, {Nonino}, {Pascale}, {Petric},
  {Pierel}, {Polletta}, {R{\"o}ttgering}, {Rutkowski}, {Smail}, {Straughn},
  {Strolger}, {Swirbul}, {Trussler}, {Wang}, {Welch}, {B. Wyithe}, {Yun},
  {Zackrisson}, {Zhang}, \& {Zhao}}]{Windhorst2023}
{Windhorst}, R.~A., {Cohen}, S.~H., {Jansen}, R.~A., {et~al.} 2023, \aj, 165,
  13, \dodoi{10.3847/1538-3881/aca163}

\bibitem[{{Witstok} {et~al.}(2022){Witstok}, {Smit}, {Maiolino}, {Kumari},
  {Aravena}, {Boogaard}, {Bouwens}, {Carniani}, {Hodge}, {Jones}, {Stefanon},
  {van der Werf}, \& {Schouws}}]{Witstok2022}
{Witstok}, J., {Smit}, R., {Maiolino}, R., {et~al.} 2022, \mnras, 515, 1751,
  \dodoi{10.1093/mnras/stac1905}

\bibitem[{{Wolfire} {et~al.}(2022){Wolfire}, {Vallini}, \&
  {Chevance}}]{Wolfire2022}
{Wolfire}, M.~G., {Vallini}, L., \& {Chevance}, M. 2022, \araa, 60, 247,
  \dodoi{10.1146/annurev-astro-052920-010254}

\bibitem[{{Zavala} {et~al.}(2021){Zavala}, {Casey}, {Manning}, {Aravena},
  {Bethermin}, {Caputi}, {Clements}, {Cunha}, {Drew}, {Finkelstein},
  {Fujimoto}, {Hayward}, {Hodge}, {Kartaltepe}, {Knudsen}, {Koekemoer}, {Long},
  {Magdis}, {Man}, {Popping}, {Sanders}, {Scoville}, {Sheth}, {Staguhn},
  {Toft}, {Treister}, {Vieira}, \& {Yun}}]{Zavala2021}
{Zavala}, J.~A., {Casey}, C.~M., {Manning}, S.~M., {et~al.} 2021, \apj, 909,
  165, \dodoi{10.3847/1538-4357/abdb27}

\bibitem[{{Zavala} {et~al.}(2024){Zavala}, {Castellano}, {Akins}, {Bakx},
  {Burgarella}, {Casey}, {Ch{\~A}{\textexclamdown}vez Ortiz}, {Dickinson},
  {Finkelstein}, {Mitsuhashi}, {Nakajima},
  {P{\~A}{\textcopyright}rez-Gonz{\~A}{\textexclamdown}lez}, {Arrabal Haro},
  {Bergamini}, {Buat}, {Backhaus}, {Calabr{\~A}{\texttwosuperior}}, {Cleri},
  {Fern{\~A}{\textexclamdown}ndez-Arenas}, {Fontana}, {Franco}, {Grillo},
  {Giavalisco}, {Grogin}, {Hathi}, {Hirschmann}, {Ikeda}, {Jung}, {Kartaltepe},
  {Koekemoer}, {Larson}, {McKinney}, {Papovich}, {Rosati}, {Saito}, {Santini},
  {Terlevich}, {Terlevich}, {Treu}, \& {Yung}}]{Zavala2024}
{Zavala}, J.~A., {Castellano}, M., {Akins}, H.~B., {et~al.} 2024, Nature
  Astronomy, \dodoi{10.1038/s41550-024-02397-3}

\end{thebibliography}
\bibliographystyle{aasjournal}



\end{document}